\documentclass{nature}

\usepackage{comment}

\usepackage{amsmath}
\usepackage[dvipsnames]{xcolor}
\usepackage{soul}
\usepackage{graphicx}
\usepackage{amssymb}
\usepackage{pifont}
\usepackage{longtable}
\usepackage{multirow}
\usepackage{arydshln}
\usepackage{hyperref}

\usepackage[switch, modulo]{lineno}
\modulolinenumbers[1]

\newcommand{\kepler}{{\it Kepler}}
\newcommand{\tess}{{\it TESS}}

\newcommand{\gaia}{{\it Gaia}}
\newcommand{\twomass}{{\it 2MASS}}

\newcommand{\pdf}{\mathrm{Pr}}

\newcommand{\cofiam}{{\tt cofiam}}
\newcommand{\polyam}{{\tt polyam}}
\newcommand{\local}{{\tt local}}
\newcommand{\gp}{{\tt gp}}

\newcommand{\multi}{{\tt MultiNest}}
\newcommand{\mann}{{\tt M\_-M\_K-}}

\newcommand{\forecaster}{{\tt forecaster}}

\newcommand{\farcs}{\mbox{\ensuremath{\,\,\!\!^{\prime\prime}}}}%

\newcommand{\wwwcooldata}{\href{https://zenodo.org/records/12734880?token=eyJhbGciOiJIUzUxMiJ9.eyJpZCI6IjFiMmVkZDI5LWFkYTItNDcxZi04ZWQxLTc5YjVlOGRhMjAxZCIsImRhdGEiOnt9LCJyYW5kb20iOiI4N2FkNjhiOTY4ZWE2NDZhYjY4MDcwZGYwNDkwNTQwMyJ9.uYFmnezpqf343z9pSfp6M-D0ZdyYmwujJ_E409KAw00E9P5hAGGwqmtyBB_Bb6wNux5GwrrxPFzaibAWL1jMsg}{this URL}}
\newcommand{\wwwcoolcode}{\href{https://github.com/davidkipping/MR_for_MKs}{this URL}}

\title{Near-circular orbits for planets around M/K-type stars with Earth-like sizes and instellations}

\author{David~Kipping$^{1}\star$,
Diana~Solano-Oropeza$^{2}$,
Daniel~A.~Yahalomi$^{1}$,\\
Madison~Li$^{1}$,
Avishi~Poddar$^{3}$,
Xunhe~Zhang$^{1}$
}

\begin{document}

\maketitle

\begin{affiliations}
 \item Dept. of Astronomy, Columbia University, 550 W 120th St., New York, NY 10027, USA
 \item Dept. of Astronomy, Cornell University, Space Sciences Building 404, Ithaca, NY 14850, USA
 \item Dept. of Astronomy, Harvard University, 60 Garden St., Cambridge, MA 02138, USA, USA
\end{affiliations}

\begin{abstract}
Recent advances have enabled the discovery of a population of potentially
Earth-like planets, yet their orbital eccentricity, which governs their climate
and provides clues about their origin and dynamical history, is still largely
unconstrained. We identify a sample of 17 transiting exoplanets
around late-type stars with similar radii and irradiation to that of Earth and
use the ``photoeccentric effect'' - which exploits transit durations - to infer
their eccentricity distribution via hierarchical Bayesian modelling. Our
analysis establishes that these worlds further resemble Earth in that their
eccentricities are nearly circular (mean eccentricity
$=0.060_{-0.028}^{+0.040}$ and $\leq0.15$), with the exception of one outlier
of moderate eccentricity. The results hint at a subset population of
dynamically warmer Earths, but this requires a larger sample to statistically
confirm. The planets in our sample are thus largely subject to minimal
eccentricity-induced seasonal variability and are consistent with emerging via
smooth disk migration rather than violent planet-planet scattering. 

\end{abstract}

\newpage

The study of exoplanets around main sequence stars began with
hot-Jupiters\cite{mayor1995}, planets that bear little resemble to Earth. This
was of course largely due to the considerable detection bias such worlds
enjoy\cite{sandford2016}, but gradually detections have now extended to more
temperate conditions, and planets with smaller radii. Despite this
progress, true ``Earth analogs'' have yet to be unambiguously identified.
Indeed such worlds are slippery to even define\cite{tasker2017}, with a
dizzying number of possible criteria such as internal structure, atmospheric
composition and dynamical environment. Most of those properties are simply
unobservable for small planets, at least thus far, and thus the best we can do
is catalog those worlds with similar radii and instellations to Earth -
quantities we can often reliably measure. As a way of concisely referring to
such planets later, we label these as ``Earth proxies''. Despite
their distinct definition, this sample provides our first glimpses about Earth
analogs in the cosmos, for the true analogs must be a subset of this broader
sample.

Earth proxies orbiting smaller stars than our Sun, M-dwarfs, have been of
particular astronomical interest. The most basic reason for this is that they
are the most observationally accessible tip of the Earth proxy population,
since their lower masses lead to greater stellar reflex motion (e.g. Proxima
Centauri\cite{anglada2016}) and their smaller radii lead to larger
shadows cast by an eclipsing (transiting) planet of any given size (e.g.
TRAPPIST-1\cite{gillon2017}). Yet more, M-dwarfs may dominate the habitable
real estate of the Universe, being the most common star type in the
cosmos, harbouring an abundance of Earth proxies and are expected to have
lifetimes far exceeding that of our Sun\cite{redsky2021}. With JWST
prioritising the observation of M-dwarf Earth proxies like TRAPPIST-1, their
status as our first window unto other Earths will surely grow in the coming
decade.

Several basic questions about this population remain, including the dynamical
environment within which they typically reside. Whilst TRAPPIST-1 has been
well-characterised in this regard\cite{agol2021}, it is the exception rather
than the rule and primarily a product of TRAPPIST-1's uncommon resonance
chain configuration. Indeed, in many cases we only know of a single planet in
the system so far, let alone whether dynamical resonances persist. The
orbital eccentricity of such worlds is arguably one of the most basic dynamical
quantities we can consider, representing a kind of dynamical temperature
of an orbit\cite{tremaine2015}, sculpting the planetary climate\cite{wang2014},
providing a vital clue to the system's history\cite{creswell2007,ford2008}
and affecting the scheduling of occultation (emission spectra)
measurements\cite{irwin1952}. Despite this, there is no previous measurement of
the eccentricity distribution for Earth proxies.

To address this, we here use asterodensity profiling\cite{MAP2012,dawson2012}
to measure the eccentricity of all known transiting Earth proxies orbiting
late-type stars, and then infer the ensemble eccentricity distribution through
hierarchical Bayesian modelling\cite{hogg2010} correcting for geometric
bias\cite{eprior2014}.

\section*{Results}
\label{sec:results}

We curated a list of all known transiting Earth proxies around
late-type stars from the NASA Exoplanet Archive\cite{akeson2013}. We filtered
for candidate exoplanets with a reported radius within a factor of two of
Earth's, an instellation within four times that of Earth and a stellar host
less than $0.75$ Solar masses (see Methods). In total, 94 candidate exoplanets
were listed in this range, of which 68 were detected with \kepler\ and 26 using
\tess.

To ensure a reliable statistical sample, it was necessary to verify that
the reported parameters were supported by an independent analysis. For the
stellar properties, the associated stellar masses ($M_{\star}$) were re-derived
using the Mann empirical relation\cite{mann2017}, which depends only upon the
parallax (here we use \gaia\ DR3\cite{gaia2016,gaia2022}) and the \twomass\
$K_S$-band apparent magnitudes\cite{twomass} (see Methods). Parallax (distance)
and magnitude yield luminosity, and since the evolutionary tracks of late-type
stars are very narrow, this leads to the remarkable outcome that we can infer
stellar masses to a few percent precision decoupled from theoretical
dependencies. The relation returns a stellar mass posterior, for which we
require that ${>}50$\% of the samples fall within Mann's calibrated range to
accept as a viable object, which removed 20 targets (3 of which had no reported
parallaxes).

Amongst the survivors, we next verify that the planet's properties can be
independently verified to satisfy our Earth proxy criteria. The photometric
light curves of each were obtained from the mission archives, detrended using
a suite of independent methods to ensure robustness\cite{teachey2018},
regressed\cite{feroz2009} using a transit light curve model\cite{mandel2002}
that accounts for known blends and then the resulting joint posteriors were
checked against our Earth proxy definition (see Methods). Cases where ${>}50$\%
of the posteriors corresponded to grazing geometries were excluded (21 targets)
since these yield degenerate eccentricity constraints. Conditional upon a
non-grazing geometry, we require that ${>}50$\% of the joint posterior samples
satisfied our criteria else the planet was rejected; a cut which removed a
further 33 targets.

Finally, we removed any planetary candidates with reported transit timing
variations (TTVs) since these manifest a confounding ``phototiming''
effect\cite{AP2014}. Since no complete catalogs exist for \tess\ TTVs,
we elected to derive our own TTVs here for this check (see Methods). Together,
3 additional planets were removed. After all three filters, just 17 Earth
proxies remain (see Table~\ref{tab:eindivs}). Of these, 7 are known to be in
multi-planet systems, and likely many of the others are too despite the fact
their companions have not been detected yet\cite{sandford2019}, hence we don't
consider it useful to attempt dividing our limited sample up in this way.
We note that amongst our sample of 17, nine are validated planets and none of the
remainder have suspiciously high false positive probabilities.


Our transit fits infer the transit duration and ingress/egress slopes (amongst
other quantities). Together, these allow us to determine the planet's orbital
speed\cite{seager2003}. The light curve model initially assumes a circular
orbit for the planet ($e=0$), and thus the orbital speed directly corresponds
to a specific semi-major axis via Kepler's Laws; although in practice it's
semi-major axis divided by stellar radius ($a/R_{\star}$) due to the scale-free
nature of a transit model. Using Kepler's Laws again, we can convert this term
into a mean stellar density, $\rho_{\star}$, by combining with orbital period.
Crucially, that's a stellar density assuming a circular orbit and it may be
compared to an independently measured density value to assess the validity of
the circular assumption - and ultimately use any deviations to infer 
eccentricity. The act of comparing densities in this way is known as
asterodensity profiling\cite{MAP2012}, and when used to infer eccentricities
it's often dubbed the photoeccentric effect\cite{dawson2012}.

\begin{figure}
\centering
\includegraphics[angle=0, width=16.0cm]{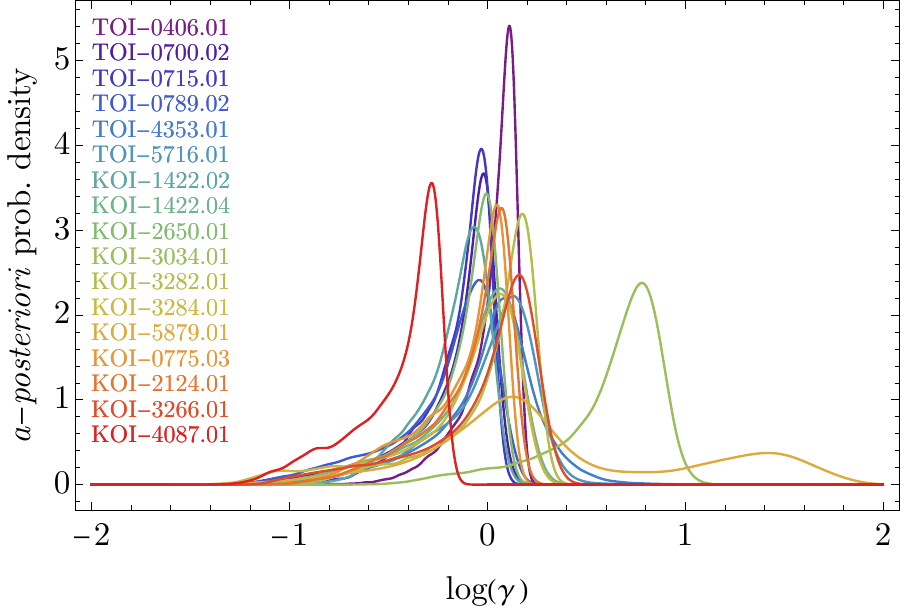}
\caption{\label{fig:gposts}
\textbf{The \textit{a-posteriori} distribution for $\log\gamma$ for 17 Earth
proxies.}
Here, $\gamma$ is an observable which tracks how eccentric an exoplanet's orbit
is, here $\log\gamma=0$ corresponds to a circular orbit measured to infinite
precision. Each exoplanet is given a unique colour, labelled on the left.
Immediately, one can see that the Earth proxy population favours near-circular
orbits.
}
\end{figure}

For each of the 17 Earth proxies, we evaluate a marginalised posterior
distribution for $\gamma$, defined as the ratio of the transit-derived stellar
density to that using the Mann empirical relation. The resulting distributions
are shown in Figure~\ref{fig:gposts}. Distributions peaking at $\gamma=1$ imply
the orbit is consistent with being circular, whereas deviations either side
imply eccentricity. It is noted that the Mann empirical relation does not
natively determine $\rho_{\star}$, just $M_{\star}$. In the Methods section, we
describe a new probabilistic, empirical relation we developed to make this
conversion using the same training sample as the original paper\cite{mann2017}.

The individual $\gamma$ measurements can be converted into individual
eccentricity measurements, although this is challenged by the fact that both
eccentricity and argument of periapsis affect $\gamma$ - leading to a
degeneracy\cite{dawson2012}. Nevertheless, one might adopt this approach and
then bin together the resulting mean/median/etc eccentricities to arrive at the
population's distribution. This approach is flawed since it exploits lossy
summary statistics, rather than the full posterior distribution; a rigorous
approach folds in all this information which can be accomplished through
hierarchical Bayesian modelling\cite{hogg2010} (HBM). In an HBM, each planet's
eccentricity is assumed to belong to some global distribution, which acts as a
prior. All planetary eccentricities (and arguments of periapsis) are inferred
jointly, as well as the shape of the population prior itself. In this way,
individual eccentricities are typically inferred with improved precision
(known as ``shrinkage'') and one obtains the global distribution from the
prior.

One disadvantage of HBMs is that the mathematical form of the population level
eccentricity distribution (the prior) must be hardcoded, even though its shape
parameters are varied. For example, one might adopt a Gaussian distribution
formulation for eccentricities, and thus fit for the population mean and
variance. \textit{A-priori}, we do not know what the correct mathematical form
of the eccentricity distribution should be. To tackle this, we attempt several
different formalisms suggested in the literature and compare their
performance.

First, we consider the classic assumption of a Rayleigh distribution, generally
expected as the outcome of planet-planet scattering events\cite{ford2008}.
Next, we consider a modification that is motivated from a statistical mechanics
argument presented by Tremaine, which defines a ``dynamical temperature'' of
said distribution\cite{tremaine2015}. Third, we try an exponential distribution
which has been previously employed for short-period exoplanets\cite{wang2011}.
These three distributions are all parameterised by a single shape term, and so
we next try a more flexible two-parameter Beta distribution, which is widely
used in exoplanet studies\cite{beta2013}. Of these four models, only the Beta
is defined over the interval $[0,1]$ as expected for eccentricity, and thus
we truncate the other three distributions and normalise appropriately.

For an eccentric orbit of semi-major axis $a$, there are more phase positions
where the planet-star separation is less than $a$ than greater, and this
leads to an enhancement in the transit probability\cite{burke2008} (which
chiefly depends upon separation). Accordingly, a population of transiting
planets are skewed towards a higher eccentricity due to geometric bias, and
this must be accounted for in our prior\cite{eprior2014}. After doing so, we
fit the HBM population parameters with priors which are uniform in
mean\cite{gelman2013} (and variance for Beta) eccentricity with a Markov Chain
Monte Carlo (MCMC) regression using 20 independent chains of 100,000 accepted
steps each. Chains were checked for mixing and convergence, followed by a
removal of burn-in chains (see Methods for details).

The four models for the eccentricity distribution were compared using the
Akaike Information Criterion\cite{akaike1974} (AIC), a model comparison tool
borne from information theory that penalises models for unnecessary complexity.
Crucially, unlike the similar Bayesian Information Criterion, the AIC does not
assume that any of the models are actually correct, but merely asks which model
best approximates the truth. This is useful here since there is no reason why
any of our competing parameterised forms would be an exact description of
reality.

\begin{figure}
\centering
\includegraphics[angle=0, width=16.0cm]{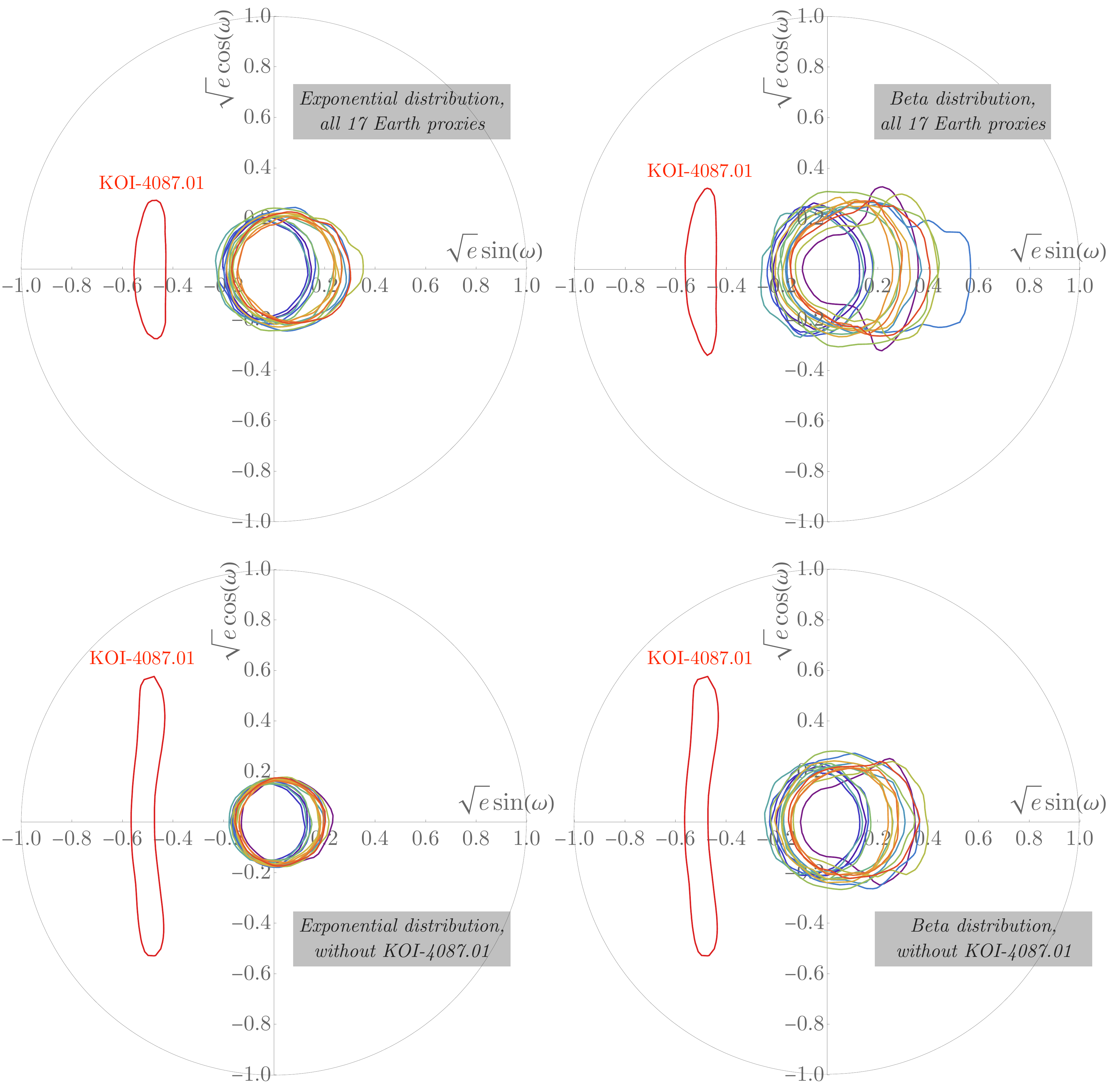}
\caption{\label{fig:indivs}
\textbf{HBM results for the individual exoplanet eccentricities.}
Each panel shows the same results but with a different underlying hierarchical
model, which is labeled inset. The contours in each panel represent the
\textit{a-posteriori} 1-$\sigma$ credible interval for the individual planets,
colour coded following Figure~\ref{fig:gposts}. The exponential model without
KOI-4087.01 is our favoured final model.
}
\end{figure}

The model comparison results are summarised in the upper panel of
Table~\ref{tab:modeltable}. The Beta distribution is nominally the favoured
model since it has the lowest AIC, but inspection of the derived individual
eccentricities reveals a clear outlier in KOI-4087.01 (see top-right panel of
Figure~\ref{fig:indivs}), which we confirm using the second-best (exponential)
model (see top-left panel of Figure~\ref{fig:indivs}). In both cases, the
ensemble distribution appears dynamically cool with the exception of this
object, thus leading us to hypothesise that KOI-4087.01 does not belong to the
same population. Indeed, one can see that KOI-4087.01's $\log\gamma$ posterior
(Figure~\ref{fig:gposts}) has effectively zero density at $0$, which defines a
circular orbit. We thus repeated our model fits for the best two models
(exponential and Beta) but excluding KOI-4087.01, which leads to improved fits
and reveals the exponential model to clearly be the favoured model 
(middle panel of Table~\ref{tab:modeltable}).

The favoured exponential distribution is characterised by a single shape
parameter, $\lambda$, which equals the reciprocal of the mean eccentricity of a
non-truncated exponential distribution. With our inference (excluding
KOI-4087.01) obtaining $1/\lambda = 0.060_{-0.028}^{+0.040}$ (see bottom-right
panel of Figure~\ref{fig:expdist}), this establishes that the Earth proxy
population is dynamically cool.

Even with the considerable uncertainty that our limited sample yields, the
eccentricity distribution appears compact and close to zero. Our own Earth has
an orbital eccentricity of just $0.0167$, which has a $p$-value of $0.24$ when
evaluated against an exponential model with our inferred mean of $0.060$. Thus,
the Earth would not be a particularly unusual member of this population,
despite the fact that our sample is distinguished in being for late-type (not
Sun-like) stars. This establishes at a population level that Earth like
planets, which recall must be a subset of our Earth proxy population, typically
have low eccentricities for late-type stars.

Naive inspection of Figure~\ref{fig:expdist} might imply that whilst the
population is indeed dynamically cool, it may be warmer than pure circular.
Afterall, we find $1/\lambda = 0.060_{-0.028}^{+0.040}$, a ${\sim}2$-$\sigma$
offset from zero. However, one must be careful here because the $\gamma$
measurements are noisy and $1/\lambda$ is a positive-definite quantity; thus 
one should always find finite probability densities in eccentric regimes,
even if the orbits were genuinely circular. This raises the question - how much
of the excess eccentricity observed here is from measurement effect, versus
genuine dynamical excitation?

To address this, we generated a fake duplicate population of the Earth proxies
with forced circular orbits. For each exoplanet, we first took the maximum
likelihood light curve fit and subtracted it from the corresponding photometry
to obtain the residuals. We then simulated a new light curve model, designed
to be representative but with a strictly circular orbit (see Methods). The
residuals were then added back onto this circular model to create a synthetic
data set. The fake Earth proxies were then subject to the same analysis as the
real data.

\begin{figure}
\centering
\includegraphics[angle=0, width=16.0cm]{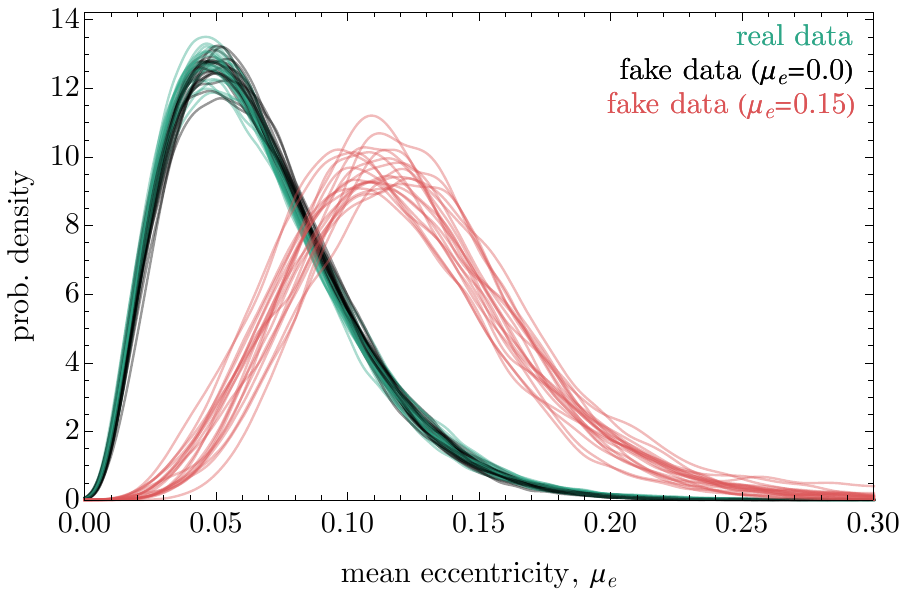}
\caption{\label{fig:comp}
\textbf{Comparison of our a-posteriori distribution for the mean eccentricity
(green) versus two injection-recoveries.}
The black lines show the inference when injecting a purely circular population,
which is indistinguishable from that obtain with the real data (green). The
red lines show the case where we inject a population where the eccentricities
follow a Rayleigh distribution with mean $\mu_e=0.15$. All cases use the same
inference model, an exponential distribution, and show 20 lines for the 20
MCMC walkers used.
}
\end{figure}

Inspecting the posteriors from the fake (circular) population, we find them to
yield very similar results (see lower panel of Table~\ref{tab:modeltable}). For
example, $1/\lambda = 0.061_{-0.028}^{+0.039}$, whereas that derived from the
real observations was $0.060_{-0.028}^{+0.040}$ (see Figure~\ref{fig:comp}).
We thus find no evidence for significant deviance between the observed
population of Earth proxies without KOI-4087.01, and a purely circular
population.

We also repeated this injection-recovery exercise for eccentric populations,
generated using Rayleigh distributions (see Methods). When fitting these with
the correct  Rayleigh model, we recover the injected mean eccentricity except
when $\mu_e\leq0.09$. Below here, the noisy data plus positive-definite nature
of eccentricity leads to a saturation behaviour towards circular-like
distributions. If we instead use a mis-specified exponential model, the mean
eccentricity is systematically offset and again saturates to circular-like
behaviour at low eccentricities ($\mu_e\leq0.15$). This suggests an upper
limit in the range of 0.09-0.15, although we have of course endevaoured to
select a model which is not grossly mis-specified via our earlier model
selection.

It is worth now returning to our one outlier, KOI-4087.01. Since it doesn't
belong with the rest, one might consider appending a second model distribution
for this object (i.e. switching to a mixture model). However, as the lone
member of this second population, it would be unconstrained and thus not a
useful exercise. Instead, we treated it separately and used uniform
eccentricity priors (but retain the geometric bias correction) to infer it has
$e=0.40_{-0.12}^{+0.20}$ and $\omega\,[^{\circ}]=271_{-38}^{+38}$.
Interestingly, this object has already been noted as likely eccentric in the
discovery paper\cite{torres2015}, where they obtained
$e\geq0.34_{-0.19}^{+0.12}$ - compatible with our solution. We note that one
outwardly surprisingly issue here is that the planet appears to transit near
apoastron, which causes $\gamma<1$, and this is geometrically less probable
than near periastron\cite{AP2014}. Given its eccentricity, we estimate that
such a planet is approximately two times more likely to be found with
$\gamma>1$ rather than $\gamma<1$, and thus its existence is hardly
implausible.

It is certainly possible then that KOI-4087.01 is simply an eccentric outlier,
but we also briefly considered alternative explanations where the measurement
is spurious. One possible explanation is that an unknown blend source
contaminates the light curve, which is consistent with the fact we measure
$\gamma<1$. Contamination would require a bright source at a separation
comparable to the pixel scale of \kepler\ (4\farcs). However, there are no
known sources\cite{gaia2016,gaia2022} found using \gaia\ within 10\farcs, nor
within 4\farcs\ using adaptive optics imaging\cite{zeigler2017}. Another
possible explanation is the phototiming effect\cite{AP2014} caused by TTVs,
but after fitting for such variations (see Methods) we find this explanation
to also be inconsistent.

At this point, we consider that the most likely explanation is that KOI-4087.01
is genuinely eccentric. It is the black swan of our sample and seemingly
belongs to a distinct and smaller population of dynamically warm Earth proxies.
How dynamically hot this subset truly is, and the overall shape of this second
distribution, cannot be reliably determined from a single example. However, if
in the future it becomes possible to extend our sample to many more objects,
the fact we obtained 1-out-of-17 belonging to a distinct group implies the hot
population comprises ${<}21.3\%$ of the ensemble to 2-$\sigma$ confidence, and
${<}34.5$\% to 3-$\sigma$ (using Binomial statistics). Using black swan
theory\cite{blackswan}, we predict that our sample would need to be extended to
34 (374) Earth proxies to have a 50\% (95.45\%) probability of seeing another 
member of this hypothesised dynamically warm population.

\section*{Discussion}
\label{sec:discussion}

We have measured the eccentricity distribution of a population of Earth proxy
exoplanets around late-type stars. During the review of this paper, another
article was published for shorter-period planets around similar stars and also
reports dynamically quiescent worlds\cite{sagear2023}. However, only 16\% of that
work's sample are Earth-proxies\footnote{Using the NEA best-reported values.}
with a stronger bias towards shorter-periods. Our principal result is that the
Earth-proxy population is dynamically cool, where our favoured model for the
ensemble is an exponential distribution of mean $0.060_{-0.028}^{+0.040}$. 
However, we identify one object which does not conform to this pattern, and may
be a member of a second dynamically warmer minority, representing $<21.3$\% of
the ensemble to 2-$\sigma$ confidence. This object, the confirmed planet
KOI-4087.01 (Kepler-440\,b), may be plausibly generated through secular
interactions with outer planets\cite{hansen2017}, which have yet to be detected
due to their long-periods.

With this object removed, the bulk population is compatible with a pure
circular-orbit population. Although only ${\sim}40$\% of our sample are
known to reside in multi-planet systems, such systems also exhibit
near-circularity\cite{vaneylen2015,xie2016} ($\mu_e = 0.064\pm0.016$) and thus
it's plausible additional undiscovered planets lurk around our target stars.
The fact that our sample is dynamically cooler than the single planet
population\cite{vaneylen2019} ($\mu_e = 0.26\pm0.05$) lends credence to this
notion, although that comparitive sample skews to higher mass stars, larger
planets and higher instellations.

An increased eccentricity naturally leads to greater swings in irradiance
and thus temperature during the orbit\cite{mendez2017}, although the climate
is heavily stabilised thanks to buffering of the thermal inertia within
oceans\cite{wang2014}. However, the location of the liquid water zone recedes
as eccentricity grows\cite{dressing2010}, which for low eccentricities is
well-described by the increase in orbit-averaged mean flux\cite{bolmont2014,
ji2022}. Accordingly, we can use our results to calculate that the outer edge
of the habitable zone shifts outwards by no more than $2$\% to 2-$\sigma$
confidence for the dominant dynamically cool population, although even this
may be an overestimate if the planets have slow rotation
periods\cite{dressing2010}. Since the dynamically warm subpopulation is a
single instance of ${\sim}0.4$ eccentricity, we are unable to estimate the
habitability consequences of its group without more representative samples.

In addition to shifting the habitable zone boundaries, moderate eccentricities
pose an additional risk, especially for M-dwarfs, since eccentricities above
0.38 are predicted to induce complete desiccation within 690\,Myr at
irradiation comparable to the modern Earth\cite{palubski2020}. Further, the
protective effect of synchronous rotation into ``eyeball'' Earth
states\cite{yang2014,kopparapu2016} only holds for circular orbits, since
eccentric planets can become trapped in spin-orbit resonances instead leading
to striped-ball climate patterns\cite{wang2014}. Our results thus support the
idea that Earth proxies rarely experience even moderate eccentricities,
considerably simplifying the target selection efforts for missions like JWST.

Compact multi-planet configurations around M-dwarfs, most notably
TRAPPIST-1\cite{gillon2017}, are thought to form through disk migration with
resonant trapping, with inner planets stalling near the disk
cavity\cite{ormel2017,unterborn2018,ogihara2022}. Such a picture naturally
gives rise to low eccentricities, even in the absence of tidal
damping\cite{tamayo2017}. However, only 7 of our 17 Earth proxies are known
to be in multi-planet systems. Our stellar sample includes both M and K
dwarfs and for K's, previous \kepler\ work reports that ${\sim}37$\% of systems
are likely single within 400\,days periodicities\cite{sandford2019}.
Therefore, our results support the idea that low eccentricities manifest for
Earth proxies in compact resonance-chain systems, but also often for planets
with no nearby and comparably-sized companions. Disk migration can thus
naturally explain the bulk of our sample, especially when one accounts for the
likelihood of numerous as-yet-undetected planets in these systems. The
largest eccentricity expected for our sample via planet-planet scattering is
${\simeq}0.30$, which is strongly excluded, although weaker scattering play a
role. An exceptional case is KOI-4087.01, for which some kind of excitation
appears necessary\cite{ford2008}.

Our work focussed on Earth proxies, planets with similar sizes and instellation
to Earth. We emphasise that these are not necessarily true Earth
analogs; they might resemble Venus for example. However, true Earth
analogs must be a subset of our Earth proxy population. Our work thus indicates
that Earth analogs around late-type stars (M/K dwarfs), stars which dominate
the cosmic population, most often have near-circular orbits similar to Earth.
This removes a potential barrier to their habitability and aids target
selection with current and future life-seeking telescopes, as well as
being broadly compatible with a quiescent and smooth disk migration origin.


\clearpage

\bibliographystyle{naturemag}
\newcounter{firstbib}

\clearpage


\begin{addendum} 
\item[Author Correspondence]
All correspondence regarding this work should be directed to D. Kipping.
\item
D.K. thanks donors Douglas Daughaday,
Elena West,
Tristan Zajonc,
Alex de Vaal,
Mark Elliott,
Stephen Lee,
Zachary Danielson,
Chad Souter,
Marcus Gillette,
Tina Jeffcoat,
Jason Rockett,
Tom Donkin,
Andrew Schoen,
Reza Ramezankhani,
Steven Marks,
Nicholas Gebben,
Mike Hedlund,
Leigh Deacon,
Ryan Provost,
Nicholas De Haan,
Emerson Garland,
Queen Rd. Fnd. Inc,
Scott Thayer,
Ieuan Williams,
Xinyu Yao,
Axel Nimmerjahn,
Brian Cartmell,
Guillaume Le Saint \&
Daniel Ohman.
D.K. acknowledges support from NASA XRP Grant \#80NSSC23K1313.
Analysis was carried out in part on the NASA Supercomputer PLEIADES (Grant \#HEC-SMD-17- 1386), provided by the NASA High-End Computing (HEC) Program through the NASA Advanced Supercomputing (NAS) Division at Ames Research Center.
This paper includes data collected by the \kepler\ and \tess\ Missions. Funding for both is provided by the NASA Science Mission directorate.
This work has made use of data from the European Space Agency (ESA) mission
\gaia\ (\href{https://www.cosmos.esa.int/gaia}{https://www.cosmos.esa.int/gaia}),
processed by the \gaia\ Data Processing and Analysis Consortium (DPAC, \href{https://www.cosmos.esa.int/web/gaia/dpac/consortium}{https://www.cosmos.esa.int/web/gaia/dpac/consortium}). Funding for the DPAC has been provided by national institutions, in particular
the institutions participating in the \gaia\ Multilateral Agreement.

\item[Author contributions] 
DK designed the project, led the analysis and interpretation, and wrote the majority of the text.
DSO managed the target list, calculated stellar posteriors and tested code across the project.
DAY. assisted with the development of method marginalisation detrending code.
ML, AP \& AZ detrended the light curves and identified bugs in earlier versions of our code.

\clearpage
\item[Author Information] Reprints and permissions information is available at www.nature.com/reprints. Correspondence and requests for materials should be addressed to DK~(email: dkipping@astro.columbia.edu).

\item[Competing Interests] The authors declare that they have no competing interests.

\end{addendum}


\begin{table*}
\caption{List of 17 Earth proxies used in our work. We list their inferred
eccentricities and arguments of periastron using our favoured HBM - an
exponential distribution but treating KOI-4087.01 as a separate class of
systems. The last column lists the inferred (not measured) $\log\gamma$ value
obtained from the HBM.
$^{\star} =$ known multi.
$^{\dagger} =$ validated/confirmed planet.
} 
\centering 
\begin{tabular}{l l l l l l l}
\hline\hline 
KOI/TOI & Other Name &
$e$ &
$\omega$ [$^{\circ}$] &
$\sqrt{e}\sin\omega$ &
$\sqrt{e}\cos\omega$ &
HBM $\log\gamma$ \\
\hline 
TOI-0406.01$^{\star}$ & & $0.044_{-0.034}^{+0.077}$ & $42_{-144}^{+91}$	& $0.06_{-0.16}^{+0.17}$ & $0.00_{-0.18}^{+0.18}$ & $0.010_{-0.026}^{+0.064}$ \\
TOI-0700.02$^{\star,\dagger}$ & TOI-700\,d & $0.031_{-0.024}^{+0.057}$ & $-19_{-114}^{+139}$	& $-0.02_{-0.14}^{+0.13}$ & $0.00_{-0.16}^{+0.17}$ & $-0.002_{-0.034}^{+0.022}$ \\
TOI-0715.01 & & $0.031_{-0.024}^{+0.058}$ & $-30_{-105}^{+142}$	& $-0.04_{-0.14}^{+0.13}$ & $0.00_{-0.16}^{+0.16}$ & $-0.004_{-0.038}^{+0.019}$ \\
TOI-0789.02$^{\star}$ & & $0.035_{-0.027}^{+0.065}$ & $-17_{-113}^{+137}$	& $-0.02_{-0.16}^{+0.14}$ & $0.00_{-0.17}^{+0.17}$ & $-0.001_{-0.039}^{+0.027}$ \\
TOI-4353.01 & & $0.045_{-0.035}^{+0.111}$ & $35_{-145}^{+93}$	&  $0.05_{-0.17}^{+0.21}$ & $0.00_{-0.18}^{+0.18}$ & $0.006_{-0.028}^{+0.086}$ \\
TOI-5716.01 & & $0.041_{-0.032}^{+0.084}$ & $29_{-142}^{+101}$	&  $0.04_{-0.16}^{+0.18}$ & $0.00_{-0.18}^{+0.18}$ & $0.005_{-0.028}^{+0.061}$ \\
KOI-1422.02$^{\star,\dagger}$ & Kepler-296\,d & $0.035_{-0.027}^{+0.066}$ & $-34_{-100}^{+143}$& $-0.04_{-0.16}^{+0.14}$ & $0.00_{-0.17}^{+0.17}$ & $-0.004_{-0.046}^{+0.022}$ \\
KOI-1422.04$^{\star,\dagger}$ & Kepler-296\,f & $0.038_{-0.029}^{+0.073}$ & $17_{-137}^{+112}$	& $0.02_{-0.16}^{+0.17}$ & $0.00_{-0.17}^{+0.17}$ & $0.002_{-0.029}^{+0.047}$ \\
KOI-2650.01$^{\star,\dagger}$ & Kepler-395\,c & $0.032_{-0.024}^{+0.058}$ & $-8_{-122}^{+132}$& $-0.01_{-0.14}^{+0.14}$ & $0.00_{-0.17}^{+0.16}$ & $0.000_{-0.032}^{+0.026}$ \\
KOI-3034.01 & & $0.043_{-0.033}^{+0.106}$ & $25_{-140}^{+102}$	& $0.03_{-0.17}^{+0.20}$ & $0.00_{-0.18}^{+0.18}$ & $0.004_{-0.030}^{+0.074}$ \\
KOI-3282.01$^{\dagger}$ & Kepler-1455\,b & $0.045_{-0.035}^{+0.096}$ & $39_{-146}^{+91}$	& $0.05_{-0.17}^{+0.19}$ & $0.00_{-0.18}^{+0.18}$ & $0.008_{-0.027}^{+0.078}$ \\
KOI-3284.01$^{\dagger}$ & Kepler-438\,b & $0.038_{-0.029}^{+0.074}$ & $17_{-136}^{+112}$ & $0.02_{-0.16}^{+0.16}$ & $0.00_{-0.17}^{+0.17}$ & $0.002_{-0.029}^{+0.047}$ \\
KOI-5879.01 & & $0.041_{-0.032}^{+0.084}$ & $18_{-136}^{+110}$ & $0.02_{-0.16}^{+0.18}$ & $0.00_{-0.18}^{+0.18}$ & $0.003_{-0.031}^{+0.055}$ \\
KOI-0775.03$^{\star,\dagger}$ & Kepler-52\,d & $0.036_{-0.027}^{+0.063}$ & $17_{-136}^{+113}$ & $0.02_{-0.15}^{+0.15}$ & $0.00_{-0.17}^{+0.17}$ & $0.003_{-0.027}^{+0.039}$ \\
KOI-2124.01 &  & $0.039_{-0.030}^{+0.069}$ & $25_{-140}^{+107}$	& $0.03_{-0.16}^{+0.16}$ & $0.00_{-0.17}^{+0.17}$ & $0.004_{-0.027}^{+0.048}$ \\
KOI-3266.01$^{\dagger}$ & Kepler-1450\,b & $0.044_{-0.034}^{+0.090}$ & $35_{-144}^{+96}$	& $0.05_{-0.17}^{+0.19}$ & $0.00_{-0.18}^{+0.18}$ & $0.007_{-0.027}^{+0.069}$ \\
KOI-4087.01$^{\dagger}$ & Kepler-440\,b & $0.400_{-0.117}^{+0.197}$ & $271_{-37}^{+38}$ & $-0.52_{-0.12}^{+0.06}$ & $0.01_{-0.41}^{+0.41}$ & $-0.296_{-0.144}^{+0.055}$ \\ [0.5ex]
\hline\hline 
\end{tabular}
\label{tab:eindivs} 
\end{table*}

\newpage

\newpage

\newpage
\begin{figure}
\centering
\includegraphics[angle=0, width=16.0cm]{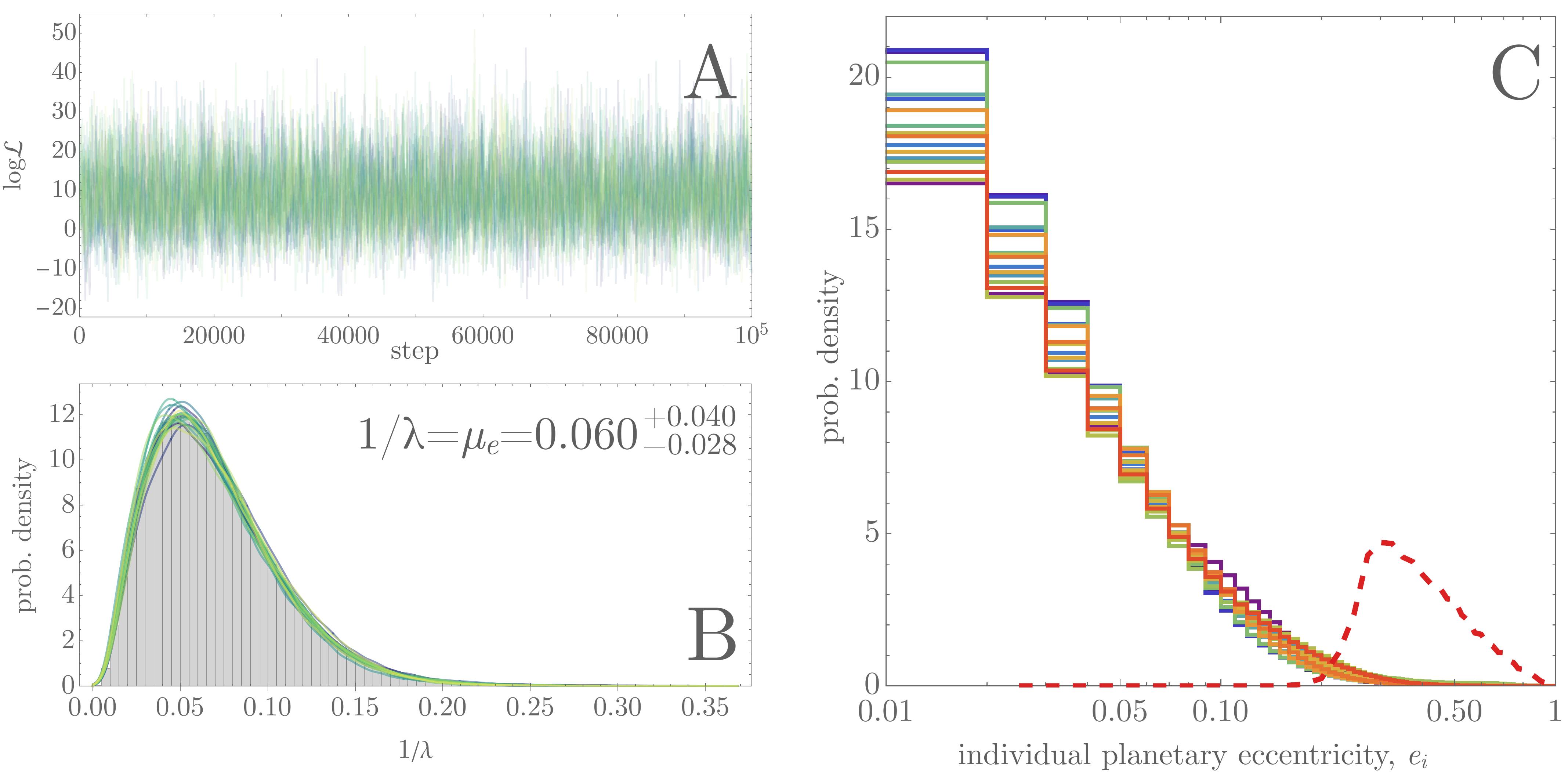}
\caption{\label{fig:expdist}
\textbf{HBM results for our favoured eccentricity distribution model (the
exponential distribution).}
A] Log-likelihoods from the 20 independent MCMC walkers (coloured blue
to yellow), showing good mixing.
B] The \textit{a-posteriori} distribution for the (only) shape
parameter controlling the eccentricity distribution, which also represents the
mean eccentricity. Grey histogram is from the combined chain, and the
colours are from the 20 independent walkers, showing excellent convergence.
C] A-posteriori distributions of the 17 individual planetary eccentricities,
following the same colour scheme as used in Figure~\ref{fig:gposts}. The dashed
one corresponds to KOI-4087.01, treated as an independent object to the
ensemble.
}
\end{figure}

\newpage

\newpage
\begin{table*}
\caption{Summary of our model testing results.} 
\centering 
\begin{tabular}{l l l l l}
\hline\hline 
Model &
$\log\hat{\mathcal{L}}$ &
AIC &
$\mu_e$ & $\sigma_e^2$ \\
\hline 
Rayleigh Dist.		& $16.05$ & $38.91$ & $0.138_{-0.034}^{+0.043}$ & $0.0052_{-0.0022}^{+0.0037}$ \\
Tremaine Dist.		& $17.76$ & $34.48$ & $0.219_{-0.052}^{+0.062}$ & $0.0231_{-0.0094}^{+0.0124}$ \\
Exponential Dist.	& $27.02$ & $15.95$ & $0.100_{-0.031}^{+0.040}$ & $0.0099_{-0.0053}^{+0.0097}$ \\
Beta Dist.			& $33.69$ & $4.63$  & $0.141_{-0.041}^{+0.050}$ & $0.0165_{-0.0080}^{+0.0119}$ \\ [0.5ex]
\hline
\textit{No KOI-4087.01} \\
Exponential Dist.	& $59.88$ & $-53.75$ & $0.060_{-0.028}^{+0.040}$ & $0.0036_{-0.0026}^{+0.0065}$ \\
Beta Dist.			& $38.44$ & $-8.88$  & $0.116_{-0.039}^{+0.049}$ & $0.0121_{-0.0070}^{+0.0119}$ \\ [0.5ex]
\hline
\textit{No KOI-4087.01, Fake $\mathbf{e}=0$} \\
Exponential Dist.	& $59.47$ & $-52.95$ & $0.061_{-0.028}^{+0.039}$ & $0.0037_{-0.0026}^{+0.0063}$ \\ [0.5ex]
\hline\hline 
\end{tabular}
\label{tab:modeltable} 
\end{table*}



\clearpage

\begin{methods}

\subsection{Target selection.}
\label{sub:targets}

The focus of this work is the population of transiting exoplanets observed by
\kepler\ and \tess\ that most closely resemble the Earth. The most
comprehensive catalog of of exoplanet candidates comes from the NASA Exoplanet
Archive (NEA\cite{akeson2013}) and thus we began by downloading this catalog at
the start of this investigation (July 7th 2022). Specifically, we
downloaded both the \kepler\ and \tess\ planetary candidates lists. Not all
\kepler\ Objects of Interest (KOIs) in the catalog are viable planet candidates
though, and so we applied a cut to remove any objects which have been
dispositioned as a likely ``FALSE POSITIVE'' by the NEA.

The definition of an ``Earth proxy'' for the purposes of this work is a planet
with a maximum \textit{a-posteriori} radius within a factor of 2 of that Earth,
and a maximum \textit{a-posteriori} instellation within a factor of 4,
similar to that used in previous studies\cite{petigura2013}. Since our work
leverages the Mann empirical mass-magnitude relation\cite{mann2017}, only
stars with masses below $0.7$\,$M_{\odot}$ are suitable. In practice, we used
an initial NEA cut of $0.75$\,$M_{\odot}$ to account for the reported
uncertainties. In total, 94 candidate exoplanets were listed in this range, of
which 68 were detected with \kepler\ and 26 using \tess.

The 68 KOIs were
KOI-253.02,
KOI-314.02,
KOI-438.02,
KOI-463.01,
KOI-494.01,
KOI-571.04,
KOI-701.03,
KOI-701.04,
KOI-719.03,
KOI-775.03,
KOI-812.02,
KOI-812.03,
KOI-817.01,
KOI-854.01,
KOI-886.03,
KOI-899.03,
KOI-904.03,
KOI-947.01,
KOI-1078.03,
KOI-1298.02,
KOI-1404.02,
KOI-1422.02,
KOI-1422.04,
KOI-1422.05,
KOI-1596.02,
KOI-1895.03,
KOI-2124.01,
KOI-2401.01, 
KOI-2418.01,
KOI-2525.01,
KOI-2626.01, 
KOI-2650.01,
KOI-2704.01,
KOI-2862.01, 
KOI-3010.01, 
KOI-3034.01,
KOI-3138.01,
KOI-3255.01, 
KOI-3266.01,
KOI-3282.01,
KOI-3284.01,
KOI-3391.01, 
KOI-3497.01, 
KOI-4087.01,
KOI-4290.01,
KOI-4356.01,
KOI-4622.01,
KOI-4742.01,
KOI-4926.01,
KOI-5009.01, 
KOI-5499.01, 
KOI-5651.01, 
KOI-5652.01, 
KOI-5789.01,
KOI-5870.01,
KOI-5878.01, 
KOI-5879.01,
KOI-5896.01, 
KOI-7099.01, 
KOI-7591.01,
KOI-7617.01,
KOI-7706.01,
KOI-7882.01,
KOI-7930.01, 
KOI-7932.01, 
KOI-8012.01,
KOI-8077.01 \& 
KOI-8174.01. 
The 26 TOIs were
TOI-175.02,
TOI-237.01,
TOI-406.01,
TOI-700.02,
TOI-700.04,
TOI-712.04,
TOI-715.01,
TOI-789.02,
TOI-1266.02,
TOI-1452.01,
TOI-2079.02,
TOI-2093.01,
TOI-2094.01,
TOI-2095.01,
TOI-2095.02,
TOI-2263.01, 
TOI-2285.01,
TOI-2296.01, 
TOI-2322.01,
TOI-2433.01, 
TOI-4328.01,
TOI-4353.01,
TOI-5388.01,
TOI-5713.01,
TOI-5716.01 \&
TOI-5728.01.

\subsection{Stellar inference.}
\label{sub:starinference}

In this work, we use the empirical mass-magnitude Mann\cite{mann2017} relation
to calculate a homogeneous and precise set of stellar masses. Unlike previous
stellar parameter estimates (e.g. those released as part of \kepler\ Data
Release 25\cite{mathur2017}), our parameters have no dependency on theoretical
uncertainties. This is a significant advantage since parameter determinations
for late-type stars are often impaired by the stars' complex spectra and
dissimilarly from the Sun. Indeed, even the best agreement models have been 
shown to systematically under-estimate stellar radii by $4.6$\%\cite{mann2015}.

However, one downside of the empirical relations is that they work best for
late-type stars (meaning we cannot include Sun-like stars), since such stars
are constrained to be dwarfs simply by the age of the Universe. Usually, a
luminosity and temperature indicator would be needed to pin a star's location
on the Hertzsprung-Russell diagram, but the fact stars of masses $\lesssim
0.7$\,$M_{\odot}$ cannot have left the main sequence means that a luminosity
indicator alone is necessary to precisely place them. In this way, the Mann
relation only needs a luminosity indicator, which in this case comes from a
\twomass\ $K_S$-band magnitude and a \gaia\ parallax. To calculate masses,
we use the \mann\ software released alongside the original
paper\cite{mann2017}, using \gaia\ DR3 parallaxes\cite{gaia2016,gaia2022} and
\twomass\ magnitudes\cite{twomass} (further details provided in
the next subsection).

Unfortunately, the Mann relation only returns a posterior distribution for
the stellar mass, whereas our work requires a stellar radius as well in order
to a) calculate planetary radii, and b) calculate stellar densities. To address
this, we derived a new empirical mass-radius relation trained on the Mann
empirically calibrated sample\cite{mann2015}. In total, 183 objects are listed
in their Tables~5, 6 \& 7, but only 171 of these have positive masses/radii for
and masses above $0.12$\,$M_{\odot}$, which we take as a conservative lower
limit to ensure brown dwarfs are confidently removed.

Inferring a new mass-radius relation is complicated by the fact that the
training data has uncertainties in both dimensions and should also plausibly
feature intrinsic variance in radius, $V$. This latter term is particularly
important, meaning our desired relation is \textit{probabilistic} (here
defined by a normal distribution) rather than \textit{deterministic}.
Accordingly, simple linear regression inversely weighted by the square of one
dimension's uncertainty will not correctly account for these complications. To
proceed, we use the likelihood function from Equation~(35) of Hogg's
pedagogical ``Fitting a model to data'' white paper\cite{hoggline}, which
defines a linear relation of gradient $m$ and offset $b$, and is given by

\begin{align}
\log \mathcal{L} &= - \sum_{i=1}^N \frac{1}{2} \log(\Sigma_i + V_R)
- \sum_{i=1}^N \frac{1}{2} \frac{\Delta_i^2}{2[\Sigma_i + V_R]}
\end{align}

where $\Sigma_i = \mathbf{\hat{v}}^T \mathbf{S}_i \mathbf{\hat{v}}^T$,
$\Delta_i = \mathbf{\hat{v}}^T \mathbf{Z}_i - b \cos\theta$,
$Z_i = \{M_i,R_i\}^T$ and

\begin{eqnarray}
 \mathbf{S}_i = 
  \begin{bmatrix}
    \sigma_{M,i}^2 & \varrho \sigma_{M,i} \sigma_{R,i} \\
    \rho \sigma_{M,i} \sigma_{R,i} & \sigma_{R,i}^2
  \end{bmatrix}.
\label{eqn:S}
\end{eqnarray}

In the above, $M_i$ and $R_i$ are the measured values of stellar mass and
radius respectively, and $\sigma_{M,i}$ and $\sigma_{R,i}$ are the respective
uncertainties. The $\varrho$ term represents the degree of correlation between
the two uncertainties - all of these are listed in the catalog
paper\cite{mann2015} used for training, except for $\varrho$ which we deal with
later. In our inference model, the free parameters are chosen to be $\theta$,
$b_{\perp}$ and $V$, where $b_{\perp} = b \cos\theta$ and $\theta = \tan^{-1}m$
(the angle between the line and $x$-axis).

We use \multi\ to infer the posterior distributions of these three free
parameters, using uniform priors on $\theta$ and $b_{\perp}$ and a log-uniform
for $V$. We repeated the analysis assuming measurement correlations of
$\varrho=0$ (idealised case) and $\varrho=1$ (worst case). We then evaluated the
predicted radii of our training set and compared to the reported radii for the
two cases. We accomplish this by generating a large number of mass samples
using the reported $M_i$ and $\sigma_{M,i}$ values, then feeding each sample
into the trained $M$-$R$ relation to generate a probabilistic radius
prediction, which itself draws input parameters from our model's joint
posterior. In this way, our predictions account for the reported uncertainty,
the model uncertainty and the intrinsic probabilistic nature of the model
(similar to the implementation of \forecaster\cite{chen2017}). We make our
code available at \wwwcoolcode.

As expected, the standard deviation in predicted radii were larger when
using $\varrho=1$ (0.0336\,$R_{\odot}$) versus $\rho=0$ (0.0328\,$R_{\odot}$).
Further, the standard deviation of the differences between the training radius
and the mean predicted radius, normalised by the uncertainty was reduced when
$\varrho=1$ (dropping from $0.348$-$\sigma$ to $0.339$-$\sigma$). Third, we
find a much higher maximum likelihood from the $\varrho=1$ model, obtaining
$\log\hat{L} = 710.2$ versus $\log\hat{L} = 617.0$. Combined with the fact that
we generally expected mass and radius to be strongly correlated in inference
problems, we adopt the $\varrho=1$ model in what follows as it produces the
more robust, conservative stellar radii. The resulting model fit is shown in
Supplementary Figure~\ref{fig:mannfit}, although the probabilistic window
around those fits is not visualised in that diagram due to the challenge of
showing both the windows plus the variance in the posterior samples themselves.

\subsection{Stellar properties.}
\label{sub:starprops}

We are now ready to evaluate the stellar masses of each of our Earth proxy host
stars using the \mann\ package, and then their stellar radii using our new
probabilistic $M$-$R$ relation. Applying to our sample of 94 candidate planets,
which are associated with 89 unique sources, we first test whether the inferred
mass is compatible with the Mann \mann\ relation.

For 3 of these sources, KOI-2626, KOI-3010 and KOI-3255, no \gaia\ parallax was
available and thus we were unable to compute a stellar mass.

For 17 other sources, the \mann\ code reported that more than 50\% of
the posteriors were outside of the range calibrated by the code, in other words
that the implied stellar mass was too large. These were
KOI-2401, KOI-2862, KOI-3391, KOI-3497, KOI-5009, KOI-5499, KOI-5651, KOI-5652,
KOI-5878, KOI-5896, KOI-7099, KOI-7930, KOI-7932, KOI-8077, KOI-8174,
TOI-2263 \& TOI-2296.

For the remaining sample, we find that the pair of empirical codes return
masses with a median uncertainty of $2.6$\%, radii with $2.4$\% uncertainty
and stellar densities to $4.9$\%. Note that because we predict radii on a
row-by-row basis on the input mass posterior, correlations are correctly
propagated into the density posterior. Our final density uncertainties are
comparable to those typically obtained via asteroseismology\cite{chaplin2014}
($4.3$\%).

The \kepler\ subset of our sample has been subject to particularly intensive
study since the mission launch over a decade ago. This provides us with an
opportunity to compare our inferred stellar masses/radii to that in the
literature. To this end, we use a \kepler\ stellar catalog focussed
specifically on M-dwarfs\cite{gaidos2016}, which employs an independent method
that combines proper motions, colours and spectroscopy to infer the stellar
properties. That study only considered \kepler\ targets though, limiting our
comparison to the 68 \kepler\ targets in our sample. Of these, we were unable
to derive stellar properties for 18, since ${>}50$\% of their posterior samples
were outside of the \mann\ package's calibration range (i.e. they do not appear
to be late-type stars). Amongst the remaining sample of 50 targets, of which
there are 46 unique stars, the Gaidos study has cross-matches for 13 (KOIs-253,
463, 571, 817, 854, 866, 899, 947, 1422, 2418, 3284, 4290, 7617). Plotting our
own stellar masses against theirs (left-panel of Supplementary
Figure~\ref{fig:starcomp}), and repeating for stellar radius (right-panel of
Supplementary Figure~\ref{fig:starcomp}), we find excellent agreement between
the two and not obvious systematic differences.

As these tend to favour mid M-dwarfs, we also sought out a comparison for one
of our latest M-dwarfs, KOI-3138. This object was chosen because it has been
the subject of particularly intense interest and scrutiny after its validation
as Kepler-1649\cite{vanderburg2020}, where the authors report a mass of
$(0.198\pm0.005)$\,$M_{\odot}$, similar to our own value of
$(0.200\pm0.085)$\,$M_{\odot}$. This comparison is somewhat unfair though
since those authors also use the \mann\ code and thus instead we can
compare to earlier work that uses spectroscopy instead\cite{angelo2017}, and
obtains $(0.219\pm0.022)$\,$M_{\odot}$ - again very similar with our value.


Blends are an important concern since any extra light would inflate the
luminosity indicator and thus the inferred stellar mass. For the final 17 Earth
proxies (found after removing unsuitable planets resulting from the transit
light curves fits, see \ref{sub:fits}), we checked each for the possibility of
a blend source. 2MASS has an angular resolution\cite{twomass} of around
3-4\farcs\ and thus we checked for evidence of companions hidden
within this range, that would otherwise corrupt the \twomass\ apparent
magnitudes.

Using \gaia\ imaging, we identify a nearby source for three of the
KOIs. KOI-2124 has a $\Delta G = -6.48$ companion 5.0\farcs\ separation,
KOI-3266 has a $\Delta G = -4.52$ companion 1.4\farcs\ separation, and KOI-7706
has two companions of magnitudes $\Delta G = -5.23$ and $-4.92$ at 4.0\farcs\
and 2.6\farcs\ separation, respectively. Converting from \gaia's visible
$G$-band magnitude to $K_S$ (a near-infrared band) is sensitive to the colour
assumed for the contaminant, although the change is unlikely to be comparable
to the range of delta magnitudes seen here. Fortunately, for all three of
these, the contaminants are not sufficiently bright to significantly influence
our inferences using \mann, for the following reasons.

A blended star will cause \twomass\ to report a brighter magnitude than
reality, and thus the reported value needs to be increased (made fainter) by
some corrective value. If two sources have a magnitude difference of
$\Delta K_S$, then the intensity ratio is $10^{0.4\Delta K_S}$ and thus the
ratio of the total intensity to the target intensity must be
$1+10^{0.4\Delta K_S}$. The magnitude of the target thus needs to be adjusted
by $+2.5\log_{10}(1+10^{0.4\Delta K_S})$. For context, the reported
uncertainties on the \twomass\ magnitudes for these three sources are 0.017,
0.028 \& 0.024 respectively. Accordingly, contaminant sources for which
$\Delta K_S < -4.50$, $-3.96$ \& $-4.13$ will lead to a correction that is
smaller than their reported uncertainties.

Given that our sample comprises late-type stars, they are necessarily
preferentially red and so it's already somewhat unlikely that an interloping
star will be redder than our targets, meaning that we expect $\Delta K_S
\lesssim \Delta G$. Indeed, we can verify this by using the $B_P-R_P$ colour
difference by \gaia\ for these contaminants. For KOI-2124, the contaminant is
indeed bluer, but for KOI-3266 no colour difference is not available.
For KOI-7706, the closer contaminant is also bluer, but the more distant
contaminant is marginally redder ($\Delta(B_P-R_P)=0.14$) - however we note
that this source is considerably fainter than our $-4.13$ threshold at
$-5.23$. As can be seen then, none of these contaminants are influencing
the \mann\ inferences at the level of measurement error, and rather than
attempt a colour conversion based on incomplete and sparse colour information,
which simply treat them as being below the noise floor in what follows.

Beyond \gaia\ imaging, companions may have also been reported in the literature
through spectroscopy or higher resolution imaging. KOI-1422 is one such source
in our sample, where Keck NIRC-2 AO imaging reveals a companion within an
arcsecond\cite{barclay2015}. At $\Delta K_S = -1.14$, this contaminant cannot
be ignored, nor is it necessary since reliable $K_S$ magnitudes exists from
direct observation (unlike with \gaia). We thus used the resolved $K_S$
magnitude of the primary in the \mann\ package in this case, in lieu of the
\twomass\ $K_S$ magnitude.

KOI-3284 also has a close companion imaged by Keck AO\cite{torres2015}, located
0.4\farcs\ away and was reported as being 2.03 magnitudes fainter in $K_S$.
Follow-up Keck observations confirm this estimate\cite{dupuy2022}, where 
averaging two measurements from July 2014 and July 2015 yield $\Delta K_S =
(2.0389\pm0.0084)$. We note that another study\cite{everett2015} arrives at a 
similar but less precise value, since the companion was detected in optical
and transformed to $K_S$ magnitude to give $\Delta K_S = 2.03_{-0.24}^{+0.19}$.
Adopting the former value, the \twomass\ reported magnitude for KOI-3284 needs
to be modified from $(11.199\pm0.020)$ to $(11.354\pm0.020)$ (note errors were
propagated in quadrature here but the $\Delta K_S$ error is simply too small to
affect things at this number of significant figures).

Similarly, a known imaged contaminant has been reported\cite{torres2017} for
KOI-0438 ($\Delta K_S = -2.160\pm0.010$),
KOI-2650 ($\Delta K_S = -7.252\pm0.079$) and 
KOI-3497 ($\Delta K_S = -1.31\pm0.47$);
and two imaged contaminants have been reported\cite{torres2017} for
KOI-0854 ($\Delta K_S = -0.299\pm0.231$ \& $-3.589\pm0.076$) and
KOI-2418 ($\Delta K_S = -2.509\pm0.062$ \& $-7.793\pm0.086$). Accounting
for these contaminants, corrects the target $K_S$ apparent magnitudes to
$11.944\pm0.030$, $12.949\pm0.030$, $10.907\pm0.109$, $13.172\pm0.101$
and $12.477\pm0.024$, respectively, which we used as our inputs for the 
\mann\ stellar inferences. The final list of inferred stellar parameters are
provided in Supplementary Table~1, along with the adopted 
effective temperatures (not inferred in this work) used later to calculate the
instellation each planet receives.


\subsection{Light curve detrending.}
\label{sub:detrending}


For each target, the light curve files were downloaded from the Mikulski
Archive for Space Space Telescopes (MAST). For \kepler, we primarily use
the long-cadence (LC) data but obtain short-cadence (SC) wherever available.
For \kepler, the data were processed as part of the 25th and final data
release issued by the \kepler\ science team\cite{thompson2018}, dubbed DR25
hereafter. For \tess, the data were similarly downloaded by MAST where
here the data is sampled consistently at 2\,minute cadence. Our work includes
\tess\ data from Sectors 1 through 48.


For all light curves, we trimmed any points with an error flag equal to
anything other than zero - thus removing points known to be afflicted by
effects such as reaction wheel zero crossings\cite{thompson2016}. Additional
outliers (e.g. unidentified cosmic ray hits\cite{morris2012}) were
removed independently for the ``Simple Aperture Photometry'' (SAP) and
``Pre-search Data Conditioning'' (PDC\cite{smith2012}) light curves, by
flagging points more than 3-$\sigma$ deviant from moving median of bandwidth 20
LC cadences.


\kepler\ and \tess\ light curves exhibit modulations in intensity due to a
variety of effects. Ultimately, the short-term modulations corresponding to a
transit are of central interest to this study, but longer term variability is
also present and introduces sizeable trends that require correction. Such
variability could originate from the instrument (e.g. focus
drift\cite{jenkins2010}) or the parent star (e.g. rotational
modulations\cite{walker2007}). In what follows, we describe our approach for
detrending these effects.

As a brief aside, we note that short-term variability on the same timescale as
the transits can also be present (e.g. pulsations in evolved
stars\cite{kallinger2016}) and is generally much more difficult to remove
since it is not separable in the frequency-domain. Consequently, attempts to
remove such noise come at grave risk of distorting a transit signal of
interest and were thus not pursued. Fortunately, such pulsations are not
expected for our sample M/K dwarfs.

For each target, we detrend the light curves of the individual transit epochs
individually, rather than imposing that the noise in one quarter/sector
need be representative of others. This is largely motivated by the fact that
the spacecrafts adjust their position and thus sources can appear on different
silicon with different optimal apertures, blend contaminations and CCD
behaviours. In addition, we adopt the approach to detrend each transit epoch
multiple different ways. The reasoning here is that, although we generally
consider each of the different methods to be fairly accurate (else we would not be
using them), we cannot guarantee that any of them is going to work in every
situation. From experience, peculiarities in particular light curves can
interact with detrending algorithms in unanticipated ways, leading to anything
from a complete failure to a subtle residual trend.

The details of the different detrending algorithms used are presented shortly,
but once in hand they are combined into a single data product (per transit
epoch) known as a ``method marginalised'' light curve\cite{teachey2018}.
In this work, we generate such light curves by simply taking the median of
the multiple detrended intensities at each time stamp. The formal uncertainty
on each photometric data point is also inflated by adding it in quadrature to
1.4286 multiplied by the median absolute deviation (MAD) between the methods.
Median statistics are used throughout here to mitigate the influence of a
failed detrending(s). In this way, we increase the robustness of our light
products against detrending choices and also inflate the errors to propagate
the uncertainty in the detrending procedure itself.

As an additional safeguard against poorly detrended light curves, we compute
two light curve statistics to measure their Gaussianity. If any of the
eight light curves fail this test, they are rejected prior to the method
marginalisation procedure. For the first test, we bin the light curves
(after removing the transits) into ever larger bins and compute the standard
deviation versus bin size, against which we then fit a linear slope in log-log
space. For such a plot, the slope should be minus one-half, reflecting the
behaviour of Poisson counting of independent measures. However, time-correlated
noise structure will lead to a shallower slope that can be used to flag such
problematic sources\cite{carter2009}. We thus generate 1000 light curves
of precisely the same time sampling and pure Gaussian noise and measure their
slopes in this way. This allows us to construct a distribution of expected
slope values. If the real slope deviates from the Monte Carlo experiments with
a $p$-value exceeding 2-$\sigma$, the light curve is flagged as non-Gaussian.

For the second test, we compute the Durbin-Watson\cite{durbin1950} statistic
of the unbinned light curves (after removing the transits). This is essentially
a test for autocorrelation at the timescale of the data's cadence, where
uncorrelated time series should yield a score of 2. As before, we test for
non-Gaussian cases by generating 1000 fake Gaussian light curves at the same
time sampling and scoring their Durbin-Watson metrics. If the real light curve
is deviant from this distribution by more than 2-$\sigma$, the light curve is
rejected.

The above describes how we combine multiple light curves detrended
independently, but we have yet to describe how these eight light curves are
generated in the first place - which we turn to in what follows. In total, four
different detrending algorithms are used, which are then to the PDC data. The
four algorithms are described in what follows.


\textbf{\cofiam:}
Cosine Filtering with Autocorrelation Minimisation (\cofiam) builds upon the
cosine filtering approach previously developed for
\textit{CoRoT}\cite{mazeh2010} data. Cosine filtering is attractive because it
behaves in a predictable manner in the frequency-domain, unlike the other
methods used here which leak power across frequency space. Fourier
decomposition of the transit morphology reveals dominant power at the timescale
of the transit duration and higher frequencies\cite{waldmann2012}. Thus, by
only removing frequencies substantially lower than this, one can ensure that
the morphology of the transit is not distorted by the process of detrending
itself. On the other hand, cosine filtering is problematic in that one could
regress a very large number of cosines to the data. Much like fitting
high order polynomials, predictions from such models become unstable at high
order. In our case, we train on the out-of-transit data (in fact the entire
quarter) and interpolate the model into the transit window, thus introducing
the possibility of high order instabilities here.

This is where our implementation deviates from that used for
\textit{CoRoT}\cite{mazeh2010}, in order to account for this effect. We detrend
the light curve up to 30 different ways, in each case choosing a different
number of cosine components to include. The simplest model is a single cosine
of frequency given by twice the baseline of available observations (thus looks
like a quadratic trend) - known as the basic frequency. At each step, we add
another cosine term of higher frequency to the function (equal to a harmonic
of the basic frequency), train the updated model, detrend the light curve and
compute statistics concerning the quality of the detrending. We continue up to
30 harmonics, or until we hit 1.5 times the reported transit duration. From the
30 options, we pick the one which leads to the most uncorrelated light curve -
as measured from the Durbin-Watson statistic evaluated on the data surrounding
(but not including) the transit (specifically to within six transit durations
either side). This local data is then exported with the data further away from
transit trimmed at this point. We direct the reader to our previous
paper\cite{hek2} for more details on this approach, including the underlying
formulae used.

\textbf{\polyam:}
Polynomial detrending with Autocorrelation Minimisation (\polyam) is similar
to the above except that the basis function is changed from a series of
harmonic cosines to polynomials. As before, 30 different possible maximum
polynomials orders are attempted from 1st- to 30th-order. And, as before,
for each epoch the least autocorrelated light curve is selected as the
accepted detrending on a transit-by-transit basis.

\textbf{\local:}
The next approach again uses polynomials, and again up to 30th order, but
this time the final accepted polynomial order is that which leads to the lowest
Bayesian Information Criterion\cite{schwarz1978} as computed on the
data directly surrounding the transit (specifically to within six transit
durations). This is arguably the simplest of the four algorithms attempted
and is a fairly typical strategy in the analysis of short-period
transiters\cite{sandford2017}.

\textbf{\gp:}
Finally, we implemented a Gaussian Process (\gp) regression to the light curve.
As with all of the methods above, the transits are masked during the regression
by using the best available ephemeris. We implemented the regression using a
squared exponential kernel where the hyper-parameters (e.g. length scale) are
optimised for each epoch independently. For consistency, we only export the
data that is within six transit durations of the transit, although technically
the entire segment ($\pm0.5$ orbital periods of each transit) is detrended.

\subsection{Light curve fits.}
\label{sub:fits}

The detrended transit light curves were modelled as an opaque spherical planet
eclipsing a spherical star with a circular symmetric intensity distribution due
to limb darkening\cite{mandel2002}.

The limb darkening of the star is modelled with a quadratic limb darkening law
using the $q_1$-$q_2$ re-parameterisation\cite{q1q2}. Since the majority of
light curves are long-cadence, the potentially significant light curve smearing
effect is accounted for by employing the numerical re-sampling
method\cite{binning2010} (with $N_{\mathrm{resamp.}}=30$). Finally,
contaminated light from nearby sources is tabulated in the {\tt fits} files as
``CROWDSAP'' and this value is used in a blend correction to each
quarter/sector's light curve using a previously published
method\cite{nightside}. We also note that the models formally assume circular
orbits with the idea being to detect eccentricity by deviations in the
inferred stellar density away from our independent values.

Regressions were executed using the multimodal nested sampling algorithm
\multi\cite{feroz2009} using 4000 live points. Typically, our model employs
seven parameters fully describe the light curve model and thus are the free
parameters in these fits. These are:
i) $P$, the orbital period of the planet
ii) $\tau$, the time of transit minimum
iii) $p$, the ratio of radii between the planet and the star
iv) $b$, the impact parameter of the planetary transit
v) $\rho_{\star}$, the mean density of the host star
vi) $q_1$, the first limb darkening coefficient
vii) $q_2$, the second limb darkening coefficient.
Uniform priors are adopted for all except $\rho_{\star,\mathrm{circ}}$
for which we use a log-uniform between $10^{-3}$\,g\,cm$^{-3}$ and
$10^{+3}$\,g\,cm$^{-3}$. For all planets, a normal likelihood function is
adopted since the data has been pre-whitened at this stage. We note for
TOI-2433.01, there was no reported NEA period and thus the object was
rejected at the stage of setting up the priors.

In some cases, known companion stars blend the transit light curve, and thus it
is necessary to account for this in our fits. Contamination factors have
already been discussed in \ref{sub:starprops}, but those were in $K_S$-band.
For the transit fits, it is necessary to convert from $K_S$-band to the band
of that used for the photometric light curve. For the companions detected
using \gaia, KOI-2124 ($\Delta G = -6.48$), KOI-3266 ($\Delta G = -4.52$) and
$G$-band ($400$-$950$\,nm) is reasonably close to the \kepler-bandpass
KOI-7706 ($\Delta G = -5.23$ \& $-4.92$), we benefit from the fact the \gaia\
($450$-$850$\,nm) and thus the \gaia\ apparent magnitudes already serve as
close approximations. The formal magnitude error on faint \gaia\ sources is 
0.029\,mag\cite{gaia2022}, but we increased this to $0.1$\,mag to account for
the bandpass difference for these three cases.

%

Besides from these, we also identified companions for seven other
stars in our sample. For KOI-3497, light curve fits are not even executed since
the stellar mass was found to be outside of the \mann\ relation. For KOI-1422,
we find two \kepler\ magnitudes reported in the literature and published nearly
coincidentally; one\cite{barclay2015} giving $\Delta K_P =
1.409_{-0.070}^{+0.085}$ and another\cite{torres2015} giving
$\Delta K_P = 1.33\pm0.22$. We elected to use the latter since it is
compatible with the first but has a larger, more conservative error.
For KOI-3284, we adopt $\Delta K_P = -1.44 \pm 0.18$ that was reported
in the same study as that used for KOI-1422.

For KOI-854 and KOI-2418, we switch to a later validation
study\cite{torres2017} by the same team, who report $\Delta K_P = -0.39\pm0.22$
and $-3.32\pm0.22$, respectively.

For KOI-438, the closest matching magnitude we can find comes from
\textit{RoboAO}\cite{baranec2017} who report $\Delta m = -3.11\pm0.04$ in a
similar band to \kepler\ (600-950\,nm). We adopt this value but again adjust
the error to $\pm0.1$ to account for slight bandpass differences. Finally, for
KOI-2650, the companion here is very faint\cite{torres2017} at $\Delta K_S =
-7.252\pm0.079$ and thus evaded detected by \textit{RoboAO}, but note the
source was previously detected using the \textit{Hubble Space
Telescope}\cite{gilliland2015} using WFC3 with the F775W filter (700-850\,nm),
obtaining $\Delta m = -7.55$, and thus we adopt $\Delta K_P = -7.55\pm0.1$.

After the fits were complete, we inspected the posteriors and eliminated
unsuitable planetary candidates. First, we remove any object for which more
than 50\% of the posterior samples imply a grazing ($b>1-R_P/R_{\star}$)
geometry, since these lead to poorly constrained stellar densities. Second,
we evaluate how many of the posterior samples satisfy our Earth proxy
criteria (a radius within twice that of Earth, an instellation within
four times that of Earth), and remove cases where ${<}50$\% of the samples
are outside of this range. Since the \mann\ relation does not calculate
effective temperatures needed for the instellation calculation, we adopted
the values from a \kepler\ catalog study\cite{berger2020} here. Our cuts
remove the majority of our objects, leaving us with just 17 Earth proxies,
11 from \kepler\ and 6 from \tess. These are the planets listed in
Table~\ref{tab:eindivs}, and for which we show the maximum likelihood
light curve fits in Supplementary Figure~\ref{fig:lcplots}. We also
present the parameters related to their suitability as an ``Earth-proxy''
in Supplementary Table~2. 

As a precaution, we checked that the grazing transit cut did not bias our
sample in terms of orbital period (and thus potentially eccentricity). A
KS-test between the periods of the grazing and non-grazing populations yields
$p=0.96$ and thus they appear consistent. Further, the median periods are very
similar at 38.3\,d and 37.1\,d, respectively. We thus find no evidence to
support that a bias was introduced as a result of this cut.

We note that the majority of our planetary candidates orbit M-dwarfs, thereby
necessitating compact orbits for Earth-like insolation levels. This raises
the possible concern that the planets are likely already circularised due
to tidal interactions with the host star. The tidal circularisation timescale
is given by (see ref. \cite{goldreich1966}):

\begin{align}
\tau_{\mathrm{circ}} &= \frac{4}{63} Q_P \Big( \frac{a^3}{G M_{\star}} \Big)^{1/2} \Big( \frac{M_P}{M_{\star}} \Big) \Big( \frac{a}{R_P} \Big)^5,
\end{align}

where $Q_P$ is the tidal quality factor of the planet. We calculated
a-posteriori samples for $\tau_{\mathrm{circ}}$ for each planetary candidate
in our sample using our posterior chains. For the planetary mass, we used the
\forecaster\ package\cite{chen2017} configured in deterministic mode.
From this, we determine median
$\log_{10}[\tau_{\mathrm{circ}}\,(\mathrm{years})]$ values ranging from
9.3 to 13.3 across our 17 objects. Only two cases have circularisation times
less than the age of the Universe, KOI-775.03 (10\,Gyr) and KOI-3284.01
(2\,Gyr). We note that field stars have a typical age of
${\sim}5$\,Gyr\cite{bonfanti2016} and thus it's likely that only one object
in our sample could be affected by tidal circularisation.

\subsection{Removing TTV systems.}
\label{sub:ttv}

Photo-eccentric effect derived eccentricities are more reliable when there is
an absence of any other asterodensity profiling effect\cite{AP2014}. To
this end, it is necessary to trim any planets exhibiting transit timing
variations (TTVs), which produces the photo-timing effect. In principle,
one could remove the photo-timing effect by correctly adjusting the
transit times for the relevant temporal perturbations. In practice, we
do not know what those perturbations are \textit{a-priori}, and simply
fitting every single epoch leads to a dynamical solution that is effectively
over-fitted when the number of transit epochs exceeds the number of
dynamical parameters (typically 7). In lieu of this, a dynamical solution
could be attempted but this would require reconstruction of all perturbing
planet transit times and is always at risk of being invalid due to the
influence of as-yet-undetected exoplanets in the system. A third solution,
the one we pursue here, is simply to remove such objects altogether.

To identify \kepler\ planets with TTVs, we can benefit from the extensive
previous analyses conducted in the literature. In particular, we remove
any KOI which was identified as a exhibiting a significant TTV system using
Tables~4 \& 5 from a \kepler\ TTV catalog analysis\cite{holczer2016}, or
indeed from a more sensitive follow-up analysis using a periodic
model\cite{ofir2016}. The result is that two KOIs were rejected,
KOI-314.02 (confirmed planet\cite{hek4}; KOI-314 c) and KOI-463.01
(validated planet\cite{morton2016}; Kepler-560 b).

To our knowledge, no comparable TTV catalog exists for \tess, with the
closest being a catalog for \tess\ hot-Jupiters\cite{ivshina2022} which is
clearly not useful here. Instead, we fit our own TTVs for the TOIs in our
sample. In our fits, we simply fix the period to the NEA reported value
and let each transit epoch have a unique transit time free parameter.
Since \multi\ dramatically slows down beyond 20 free parameters, we
split long sequences into multiple segments, similar to earlier
work\cite{hek6}. The transit times were regressed to a linear + sinusoidal
model which slid across a grid of candidate periods uniformly spaced
in frequency (i.e. a modified Lomb-Scargle periodogram). We repeated
this but for synthetic transit times generated assuming a strictly
linear ephemeris + Gaussian noise about their individual uncertainties.
The results of the latter were used to define a false-alarm probability
(FAP) curve as a function of $\Delta\chi^2$ improvement. Applying this
to real periodograms, we calculated the FAP (tailored for each TOI) of
the maximum $\Delta\chi^2$ peak period. We find that only one
of our TOIs has a peak period with a FAP lower than $0.27$\% (3-$\sigma$
in $1-$FAP) - that is TOI-2079.02. This object has a 3.6-$\sigma$
FAP indicative of a 30-minute amplitude TTV, and thus it joins
KOI-314.02 and KOI-463.01 as being excluded in our analysis.

A list of our transit times are made available at \wwwcooldata.
We do no calculate TTVs for TOI-2263.01 or TOI-2296.01
since the host star exceeds the \mann\ relation limits. We also do not
calculate TTVs for TOI-4328.01 or TOI-4353.01, since both only have
2 available epochs.

A possible concern might be that excluding three planetary candidates due to
exhibiting TTVs could bias our sample towards low eccentricity systems. This is
because the TTV amplitude scales with the conjugate of the sum of the complex
free eccentricities\cite{lithwick2012}. However, we note that TOI-2079.02 has
only seven epochs and thus can be fit in a single \multi\ regression as the
bulk of our sample was, and doing so yields $\log\gamma = 0.21_{-0.25}^{+0.15}$
and is thus also consistent with circular. Further, the TTVs of KOI-314 have
been previously studied\cite{jontof2015} who determine KOI-314.02 has an
eccentricity of ${\sim}0.068$. Thus, we find no evidence that excluding these
objects introduces a bias.

\subsection{Hierarchical Bayesian modelling.}
\label{sub:hbm}

A hierarchical Bayesian model (HBM) is a rigorous approach for deriving
population-level parameters\cite{hogg2010}. It essentially treats the shape
parameters governing the priors (hereafter ``hyper-parameters'',
$\boldsymbol{\phi}$) as unknown terms to be inferred, in addition to
the individual parameters defining each sample member, $\boldsymbol{\theta}_i$.

In our case, $\boldsymbol{\theta}_i$ represents the vector $\{e_i,\omega_i\}$,
and we further define $\boldsymbol{\Theta}$ to be the vector of these vectors
across all planets in our sample, hence it is a matrix.
We assume that $\omega_i$ is intrinsically uniformly distributed on the
sky whereas the intrinsic prior for $e_i$ takes some parametric form. The exact
form here has been argued to take numerous parametric forms\cite{ford2008,
wang2011,beta2013,tremaine2015}, which in turn depend on some shape parameters,
$\boldsymbol{\phi}$. For example, in the case of the Beta distribution prior,
$\boldsymbol{\phi}=\{\alpha,\beta\}$.

Given some data, $\mathcal{D}$, we can infer both $\boldsymbol{\theta}_i$ and
$\boldsymbol{\phi}$ together by writing out Bayes' theorem:

\begin{align}
\pdf(\boldsymbol{\theta}_i,\boldsymbol{\phi}|\mathcal{D}) &\propto
\pdf(\mathcal{D}|\boldsymbol{\theta}_i,\boldsymbol{\phi}) \pdf(\boldsymbol{\theta}_i,\boldsymbol{\phi}) 
\end{align}

which can be further expanded to

\begin{align}
\pdf(\boldsymbol{\theta}_i,\boldsymbol{\phi}|\mathcal{D}) &\propto
\pdf(\mathcal{D}|\boldsymbol{\theta}_i,\boldsymbol{\phi}) \pdf(\boldsymbol{\theta}_i|\boldsymbol{\phi}) \pdf(\boldsymbol{\phi}).
\end{align}

Here, the first term on the right-hand side (RHS) is the likelihood, the second
term is the prior and the third term is the so-called ``hyper-prior''. In our
case, we also know that the planets are transiting and thus imparts an important
bias into the inference\cite{eprior2014} that must be accounted for. We denote
this extra condition as $\hat{b}$ and thus have

\begin{align}
\pdf(\boldsymbol{\theta}_i,\boldsymbol{\phi}|\mathcal{D},\hat{b}) &\propto
\pdf(\mathcal{D}|\boldsymbol{\theta}_i,\boldsymbol{\phi}) \pdf(\boldsymbol{\theta}_i|\boldsymbol{\phi},\hat{b}) \pdf(\boldsymbol{\phi}).
\end{align}

where it can be seen that only the prior is directly affected by this bias
(i.e. certain values of $\boldsymbol{\theta}_i$ are more likely
\textit{a-priori} because of geometric bias).

In our case, the data here is taken to be the individual exoplanetary
$\log\gamma$ posteriors from our earlier transit light curve fits. This is
convenient as this one-dimensional marginalised posterior contains effectively
all of the information about eccentricity\cite{kipping2008}. To convert these
posteriors into likelihood functions, we use a smooth kernel density estimator
function using a Gaussian kernel and a bandwidth selection using Scott's
estimator method\cite{scott2015}. Once trained, we can now evaluate the
posterior density value (i.e. the likelihood) at any arbitrary choice of
$\log\gamma$ on a continuous scale. Since $\log\gamma$ directly depends upon
$e$ and $\omega$, this defines our likelihood function for a single object.

We generalise to the ensemble by simply taking the product over all objects:

\begin{align}
\pdf(\boldsymbol{\Theta},\boldsymbol{\phi}|\mathcal{D},\hat{b}) &\propto
\prod_{i=1}^N \Bigg( \pdf(\mathcal{D}|\boldsymbol{\theta}_i,\boldsymbol{\phi}) \pdf(\boldsymbol{\theta}_i|\boldsymbol{\phi},\hat{b}) \pdf(\boldsymbol{\phi}) \Bigg).
\label{eqn:HBM}
\end{align}

We calculate the transit-bias corrected joint prior for each of the proposed
analytic forms (Rayleigh, Tremaine, Exponential, Beta) following the same
recipe used to originally in the case of the Beta
distribution\cite{eprior2014}.

In total, for $N$ exoplanets, we have $2N$ free parameters describing
$\boldsymbol{\Theta}$ and $\mathrm{len}(\boldsymbol{\phi})$ parameters
describing $\boldsymbol{\phi}$. At most this is 36 free parameters, but varies
depending on the model and whether KOI-4087.01 is included or not.
A graphical model depicting our HBM structure in the example of a Beta
distribution is illustrated in Supplementary Figure~\ref{fig:graphical_hbm}.

We use a Markov Chain Monte Carlo (MCMC) Metropolis sampler to explore
the parameter space, using 20 independent chains of 100,000 accepted
steps each. Rather than walk in the native parameters described thus far,
we convert $\{e_i,\omega_i\} \to \{h_i',k_i'\} =
\{\sqrt{e_i}\sin\omega_i,\sqrt{e_i}\cos\omega_i\}$ to prevent stepping
problems near the boundary condition of a circular orbit. Further,
for the hyper-parameters, we walk in the mean of the eccentricity
for one parameter distributions, and the mean + variance for 
two parameter distributions (using uniform hyperpriors), as generally
recommended for HBMs\cite{gelman2013}.

Chains were inspected for convergence and mixing, with the top-left panel of
Figure~\ref{fig:expdist} showing an example for the exponential distribution
(without KOI-4087.01), where excellent mixing is evident. The inferred
hyper-parameters, $\boldsymbol{\phi}$, for each model are listed in
Table~\ref{tab:modeltable}, and the individual parameters in
Table~\ref{tab:eindivs}. In all cases, we remove burn-in samples defining
burn-in as the point where the likelihood first exceeds the entire chain's
median likelihood. The entire joint posterior samples for all 7 models listed
in Table~\ref{tab:modeltable} are made publicly available at \wwwcooldata.
We also show summary figures for each model, similar to
Figure~\ref{fig:expdist}, but for the other models tested in Supplementary
Figures~\ref{fig:tremplot}-\ref{fig:expnXCplot}.

\subsection{Generating fake circular planets.}
\label{sub:fakes}

In order to provide some context to our final results, we seek to repeat the
analysis but using a fake data set as input. Specifically, we generate a fake
data set where all the transiting planets in our sample are forced to be on
exactly circular orbits. The results produced from this will thus provide 
guidance as to what a truly circular population should look like.

Our approach here starts with an initial light curve fits of each exoplanet
in our sample
(as discussed earlier). From the joint posteriors, we identify the
\textit{maximum a-posteriori} (MAP) solution and calculate the corresponding
light curve. The residuals of this model against the data are then saved and
will be used shortly.

Next, we repeat our original transit fits but change the prior on the stellar
density, forcing to be fixed to the MAP solution as derived from our earlier
stellar inference.
By forcing the transit light curve
density to equal the ``true'' density, we are effectively imposing a circular
orbit. The idea here is to find a light curve shape which closely resembles
that implied by the data, but imposes the condition we seek of circularity. The
MAP solution produced by this new fit can be used as a template for our fake
planet injection.

One caveat here is that the impact parameter should not be grazing ($b>1-p$) as
such systems were explicitly rejected in our Earth proxy down selection.
Indeed, even near-grazing solutions is often sufficient to
cause the subsequent posteriors to have considerable weight in the grazing
regime. Accordingly, if $b>0.75$, we re-generated the impact parameter
by drawing from the Sandford distribution\cite{sandford2016} for $b$. Further,
we set the limb darkening parameters to be equal to their \textit{a-posteriori}
median values, rather than MAP values, since grazing solutions often push limb
darkening terms into extreme territory.

The template and the previously saved residuals were now combined to create a
representative fake, circular orbit data set. Armed with the set, we repeated
our entire inference pipeline, including the HBM, to infer the resulting
eccentricity distribution to provide context for the real data (see
Supplementary Figure~\ref{fig:expnXCplot}).

\subsection{Generating fake eccentric planets.}
\label{sub:fakes}

In addition to injecting a purely circular population, and establishing a
circular retrieval, it is instructive to inject a dynamically warm population
and ensure the retrievals don't stubbornly favour a circular ensemble.

To generate our warm populations, we need to choose a true underlying
eccentricity distribution. In what follows, we elected to assume the true
distribution was a Rayleigh (truncated over the interval $[0,1]$) whereas
we tried two inference models, a Rayleigh and an exponential. With the former,
we are thus fitting the correct model to the true generating function, whereas
the latter considers the case where the model is mis-specified but matches
the final model used with the real data.

We generated joint eccentricity and argument of periastron samples for the
Rayleigh distribution accounting for the geometric bias inherent to the transit
method, following the method described in reference\cite{eprior2014}. This
entails a normalisation via numerical integration which we found to be unstable
for $\sigma < 0.03$ and thus our injected populations span $\sigma=0.03$ to
$0.20$ in $0.01$ steps. For each injected pair of $e$-$\omega$ (one for each
planet), we computed the corresponding $\log\gamma$ value.

To create a representative noisy input for our HBM inference, we read in the
$\log\gamma$ posteriors obtained previously in the purely circular injection
case. We then add our injected $\log\gamma$ values onto these posteriors,
shifting them to account for the effect of eccentricity whilst maintaining the
same shape, variance and tailedness expected from realistic transit fits. These
perturbed $\log\gamma$ posteriors are then fed into the HBM as before, except
we reduce the number of MCMC chains to $10^4$ steps per walker (still using 20
walkers) for computational expediency. The entire procedure was repeated three
times, such that we in fact created three realisations of the injected data set
for each possible $\sigma$ value, in order to provide some robustness checks
against the random seeds used for the injections themselves.

The results of this exercise are summarised in Supplementary
Figure~\ref{fig:hotfakes}, for both the Rayleigh (left) and exponential (right)
models. Note that we append the circular injection result from the previous
subsection into these results, giving us 3x18 + 1 = 55 injection-recoveries
altogether. We compare the results in terms of $\mu_e$, since this is
well-defined for both models.

When using the correct model (Rayleigh), we find excellent agreement (within
1-$\sigma$) between the injections and retrievals down to $\mu_e{\sim}0.09$.
Below this point, the inferences flatten out and consistently over-estimate the
mean eccentricity. As seen earlier with the circular injection, we argue this
is caused by a combination of the inherent uncertainties and $\log \gamma$ and
the fact that $\mu_e$ is a positive-definite quantity. We regressed a piecewise
function to the left panel of Supplementary Figure~\ref{fig:hotfakes}, which is
designed to saturate below some threshold point, $\mu_0$, and follow the 1:1
line otherwise. Our regression yeilds $\mu_0 = 0.091$, for which the $\chi^2$
of the fit is 12.9 for 55 data points.

When using the mis-specified model (here an exponential model), we see the same
flattening effect at low eccentricities, but also a systematic offset away from
the 1:1 line beyond this point. We again regress a piecewise function to the
results, but now include a systematic offset term. This reveals the flattening
effect is in effect below $\mu_e<0.15$ and above that we have a systematic
underestimation in the mean eccentricity of $0.078$ ($\chi^2=5.8$). This
behaviour is shown in the right panel of Supplementary
Figure~\ref{fig:hotfakes}. We also include a shaded region for the 68.3\%
credible interval of $\mu_e$ derived from our real data using the exponential
model. This reveals that our real result is consistent with the saturation
regime. In other words, the eccentricity of our population once again appears
consistent with zero.

An obvious question at this point is just how much higher than zero could the
mean eccentricity be? Obviously the saturation threshold here occurs at
$\mu_0=0.15$, and thus we might reasonably conclude that our result implies a
mean eccentricity below this point. For context, we note this is about
2-$\sigma$ deviant from our best reported value ($\mu_e=0.06_{-0.03}^{+0.04}$).
However, this threshold was established for one particular set of assumptions
- where eccentricities truly follow a Rayleigh distribution but we fit them
with an exponential distribution. Yet, our analysis of the real data indicates
that a Rayleigh distribution is not favoured and we have endeavoured to select
the most favourable model. If our model is correctly specified, we might
instead conclude a threshold of $\mu_0=0.09$, based on our experiments. In
reality, we cannot know what the true distribution is, but these experiments
suggest a $\mu_0$ value between $0.09$ and $0.15$, in going from the extremes
of a best-case scenario (correctly specified model) to a worst-case scenario
(mis-specified model), respectively.

Much of this discussion speaks to the fundamental limitation with HBMs. They
always insist on some assumed parameterised form for the underlying
distribution and in general even the most favoured model could be slightly
mis-specified.

\subsection{KOI-4087.01 checks.}
\label{sub:4087}

Here, we briefly expound upon TTV test for KOI-4087.01 that was mentioned in
the main text. Transit timing variations (TTVs) occur when the planet is
gravitationally perturbed from a strictly Keplerian orbit, and thus no longer
follows a linear ephemeris. Since our light curve fits assume a linear
ephemeris, the presence of TTVs would violate that assumption and lead to
erroneous inferences\cite{AP2014}, including the $\log\gamma$ parameter used
for eccentricity determination in this work.

KOI-4087.01 is not known to exhibit TTVs, but the significant evidence for a
non-zero $\log\gamma$ and the direction of that deviance is potentially
compatible with TTVs. We thus decided to perform our own TTV fits here as a
check. The light curve model was modified such that each transit epoch was
given a unique free parameter describing its time of transit minimum. Here,
we identify 10 such epochs, all recorded in long-cadence mode. Our inferences
obtain transit times with coarse uncertainties of around 20\,minutes,
which is not surprising given the diminutive size of this exoplanet (and thus
low signal-to-noise). The TTVs are shown in Supplementary
Figure~\ref{fig:4087_ttvs}.

Running a periodogram through the times, we find no evidence for a periodic
signature, with the best sinusoid only scoring $\Delta\chi^2=4.7$. We
generated 100 random set of TTVs perturbed about a linear ephemeris by
their uncertainties to calculate the false alarm probability (FAP) and
find that this signal has a 50\% FAP and thus is clearly not significant
(see Supplementary Figure~\ref{fig:4087_ttvs}).

\subsection{Planet-planet scattering.}
\label{sub:scattering}

Our HBM analysis indicates that the Earth-proxy sample around late-type stars
is consistent with $e\lesssim0.15$, barring rare exceptions. In what follows,
we consider how compatible such a result is with the predictions of eccentricity
excitation through planet-planet scattering events. To this end, we use the
approximate maximum eccentricity prediction formula from Zhu \& Dong's recent
review\cite{zhu2023} (their Equation~17), given by

\begin{align}
e_{\mathrm{max}} \simeq \sqrt{\frac{2 M_P a}{M_{\star} R_P}}.
\end{align}

In our case, $a/R_P$ is equivalent to $a_R/p$ - for which we have posterior
samples from our transit fits. The stellar mass, $M_{\star}$ is known from our
empirical mass-radius relation but the planetary mass, $M_P$, is not directly
observed. We again turn to \forecaster\cite{chen2017} to estimate this
quantity, based on the observed radius, $R_P$. We calculated posteriors for
$e_{\mathrm{max}}$ with 1-$\sigma$ credible intervals of
$0.27_{-0.02}^{+0.01}$ (TOI-406.01),
$0.26_{-0.02}^{+0.02}$ (TOI-700.02),
$0.33_{-0.03}^{+0.02}$ (TOI-715.01),
$0.22_{-0.03}^{+0.02}$ (TOI-789.02),
$0.45_{-0.06}^{+0.03}$ (TOI-4353.01),
$0.19_{-0.03}^{+0.03}$ (TOI-5716.01),
$0.27_{-0.02}^{+0.02}$ (KOI-1422.02),
$0.38_{-0.05}^{+0.03}$ (KOI-1422.04),
$0.26_{-0.03}^{+0.02}$ (KOI-2650.01),
$0.39_{-0.05}^{+0.03}$ (KOI-3034.01),
$0.36_{-0.05}^{+0.02}$ (KOI-3282.01),
$0.26_{-0.04}^{+0.03}$ (KOI-3284.01),
$0.39_{-0.08}^{+0.37}$ (KOI-5879.01),
$0.30_{-0.03}^{+0.02}$ (KOI-775.03),
$0.25_{-0.02}^{+0.02}$ (KOI-2124.01),
$0.33_{-0.05}^{+0.03}$ (KOI-3266.01) and
$0.35_{-0.04}^{+0.02}$ (KOI-4087.01).

The median of this sample is $0.30$, which is strongly excluded by the analysis
presented in this work. We thus conclude that the scenario of planet-planet
scattering producing maximal eccentricities for Earth-proxies is excluded.
However, since this is only the maximal case, it is certainly possible and
indeed to some degree inevitable that planet-planet scattering will contribute
some finite amount of eccentricity to the sample.

\end{methods}


\clearpage
\begin{addendum}
\item[Data Availability]
This paper includes data collected by the \kepler\ and \tess\ mission, which are
publicly available from the Mikulski Archive for Space Telescopes (MAST) at the
Space Telescope Science Institute.
The data that support the plots within this paper and other findings of this
study are available at \wwwcoolcode\ and \wwwcooldata; or from the corresponding
author upon reasonable request.

\item[Code Availability]
The \multi\ regression algorithm\cite{feroz2009} is publicly available at
\href{https://github.com/farhanferoz/MultiNest}{https://github.com/farhanferoz/MultiNest}.
The \mann\ software package\cite{mann2017} is publicly available at
\href{https://github.com/awmann/M_-M_K}{https://github.com/awmann/M\_-M\_K}.
Our software package to calculate probabilistic stellar radii from mass is
publicly available at \href{https://github.com/davidkipping/MR_for_MKs}{https://github.com/davidkipping/MR\_for\_MKs}.
\end{addendum}




\newpage
\textbf{\Large{Supplementary Information}} 

\begin{sfigure}
\centering
\includegraphics[angle=0, width=16.0cm]{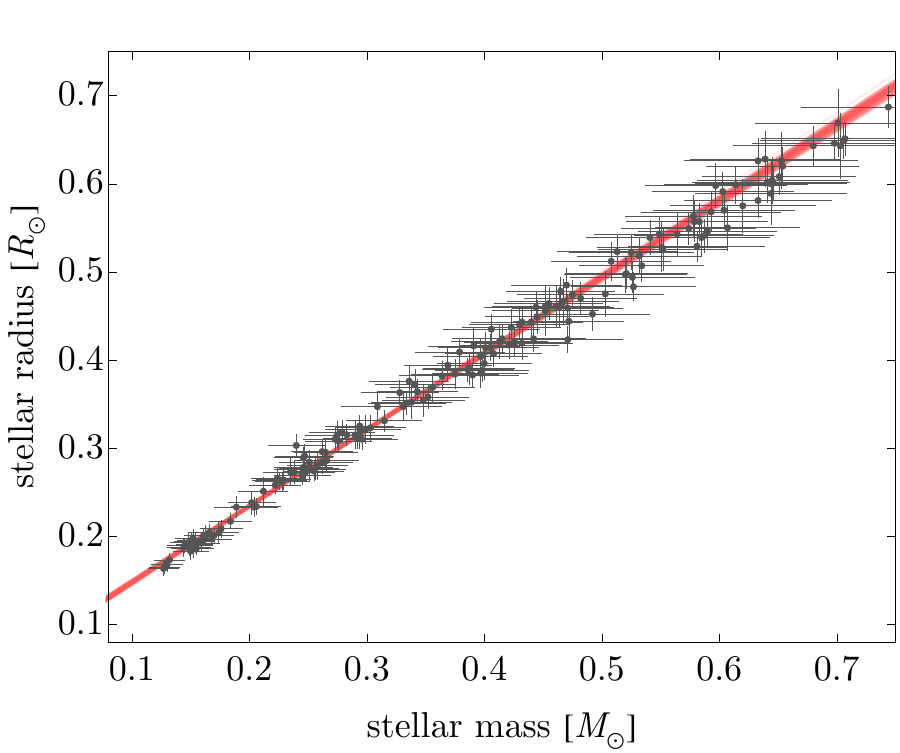}
\caption{\label{fig:mannfit}
\textbf{Training data used to infer our probabilistic mass-radius relation for
late-type stars.}
Black points are 171 late-type stars with empirically determined masses and
radii taken from previous work\cite{mann2015}. We show 100 red lines
representing fair draws from our joint posterior for the mass-radius relation.
}
\end{sfigure}

\begin{sfigure}
\centering
\includegraphics[angle=0, width=16.0cm]{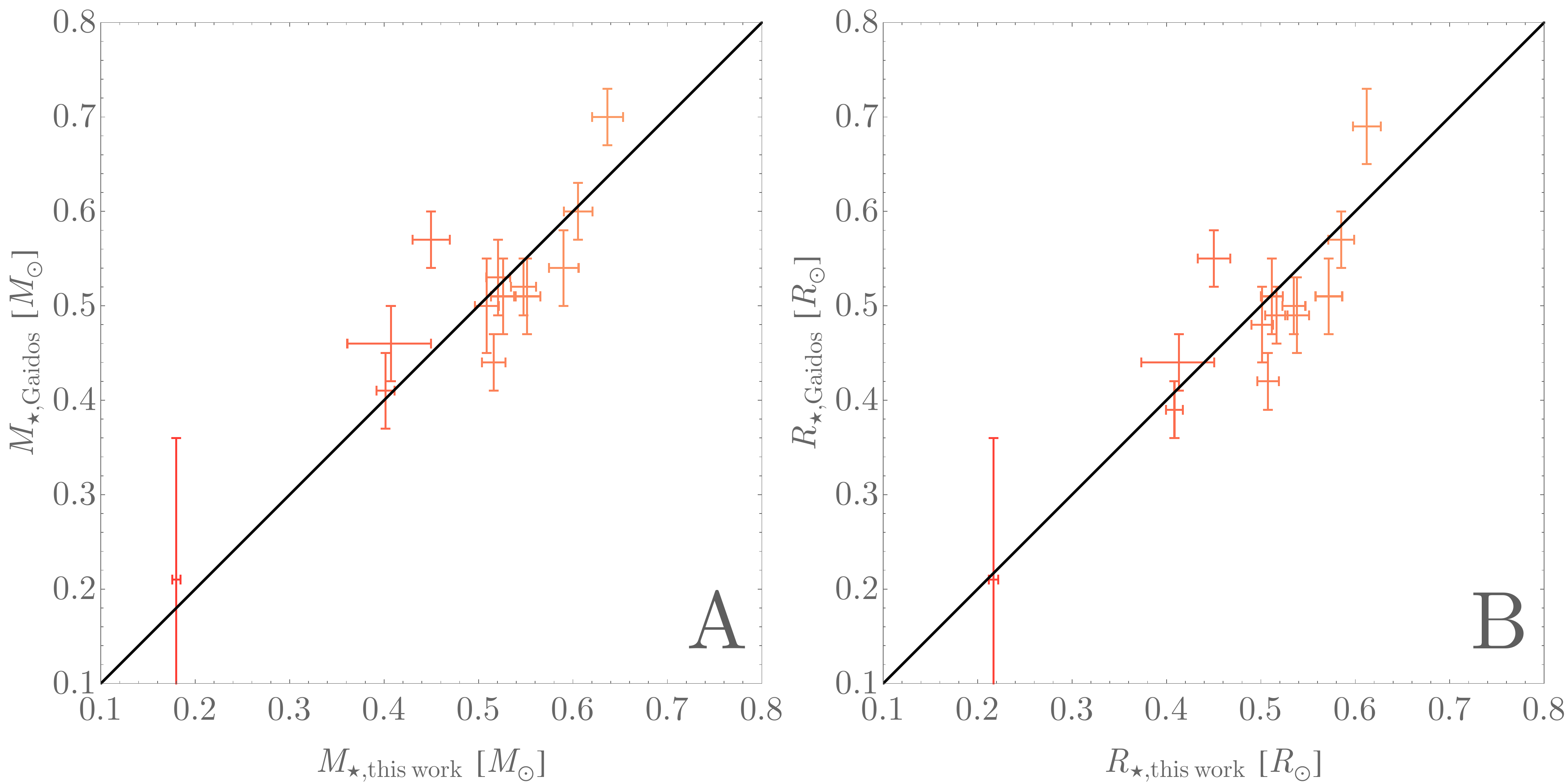}
\caption{\label{fig:starcomp}
\textbf{Comparison of stellar properties inferred for thirteen stars cross-matched
between our own sample and that of Gaidos et al. (2016).}
A] Comparison of stellar masses, showing 1-$\sigma$ credible intervals with
the black crosses and the line of agreement in black. B] Same as A,
except for stellar radii.
}
\end{sfigure}

\begin{stable}
\caption{Stellar parameters adopted. Column 2
uses the \mann\ package, column 3 our new $M$-$R$ relation.
The last column is adopted from literatures sources; for \tess\ targets we use the
TIC\cite{stassun2019}, and for \kepler\ we use the Berger
catalog\cite{berger2020}. The exceptions are KOI-1422 which comes from
\cite{barclay2015}, and TOI-700 which comes from \cite{rodriguez2020}.
} 
\centering 
\begin{tabular}{l l l l l}
\hline\hline 
KOI/TOI &
$M_{\star}$\,[$M_{\odot}$] &
$R_{\star}$\,[$R_{\odot}$] &
$\rho_{\star}$\,[$\mathrm{g}\,\mathrm{cm}^{-3}$] &
$T_{\mathrm{eff,adopted}}$\,[K] \\ [0.5ex] 
\hline 
TOI-0406 & $0.3791_{-0.0090}^{+0.0091}$ & $0.3891_{-0.0084}^{+0.0086}$ & $9.07_{-0.38}^{+0.42}$ & $3324\pm157$ \\ 
TOI-0700 & $0.3743_{-0.0093}^{+0.0094}$ & $0.3849_{-0.0087}^{+0.0088}$ & $9.25_{-0.41}^{+0.44}$ & $3461\pm66$ \\ 
TOI-0715 & $0.1981_{-0.0047}^{+0.0047}$ & $0.2324_{-0.0050}^{+0.0051}$ & $22.3_{-1.0}^{+1.1}$   & $3187\pm157$ \\ 
TOI-0789 & $0.3607_{-0.0088}^{+0.0088}$ & $0.3732_{-0.0082}^{+0.0083}$ & $9.78_{-0.42}^{+0.45}$ & $3461\pm157$ \\ 
TOI-4353 & $0.422_{-0.010}^{+0.010}$    & $0.4266_{-0.0092}^{+0.0094}$ & $7.67_{-0.33}^{+0.35}$ & $3614\pm157$ \\ 
TOI-5716 & $0.1931_{-0.0047}^{+0.0048}$ & $0.2281_{-0.0050}^{+0.0051}$ & $23.0_{-1.1}^{+1.1}$   & $3331\pm157$ \\ 
KOI-1422 & $0.407_{-0.046}^{+0.043}$    & $0.413_{-0.040}^{+0.037}$    & $8.1_{-1.2}^{+1.7}$    & $3740\pm130$ \\ 
KOI-2650 & $0.619_{-0.016}^{+0.016}$    & $0.597_{-0.015}^{+0.015}$    & $4.11_{-0.19}^{+0.21}$ & $4096\pm69$ \\ 
KOI-3034 & $0.624_{-0.016}^{+0.016}$    & $0.601_{-0.014}^{+0.014}$    & $4.05_{-0.19}^{+0.20}$ & $4175\pm59$ \\ 
KOI-3282 & $0.585_{-0.015}^{+0.015}$    & $0.568_{-0.014}^{+0.014}$    & $4.51_{-0.21}^{+0.23}$ & $4050\pm64$ \\ 
KOI-3284 & $0.590_{-0.015}^{+0.016}$    & $0.572_{-0.014}^{+0.014}$    & $4.45_{-0.21}^{+0.22}$ & $3957\pm85$ \\ 
KOI-5879 & $0.619_{-0.016}^{+0.016}$    & $0.597_{-0.014}^{+0.015}$    & $4.11_{-0.19}^{+0.20}$ & $4148\pm49$ \\ 
KOI-0775 & $0.626_{-0.016}^{+0.016}$    & $0.603_{-0.014}^{+0.014}$    & $4.03_{-0.18}^{+0.20}$ & $4164\pm62$ \\ 
KOI-2124 & $0.720_{-0.021}^{+0.021}$    & $0.684_{-0.019}^{+0.019}$    & $3.17_{-0.17}^{+0.18}$ & $4485\pm59$ \\ 
KOI-3266 & $0.691_{-0.018}^{+0.018}$    & $0.659_{-0.016}^{+0.016}$    & $3.40_{-0.16}^{+0.17}$ & $4459\pm75$ \\ 
KOI-4087 & $0.645_{-0.016}^{+0.016}$    & $0.619_{-0.014}^{+0.015}$    & $3.83_{-0.17}^{+0.19}$ & $4171\pm56$ \\ 
\hline\hline 
\end{tabular}
\label{tab:stars} 
\end{stable}


\begin{sfigure}
\centering
\includegraphics[angle=0, height=20.0cm]{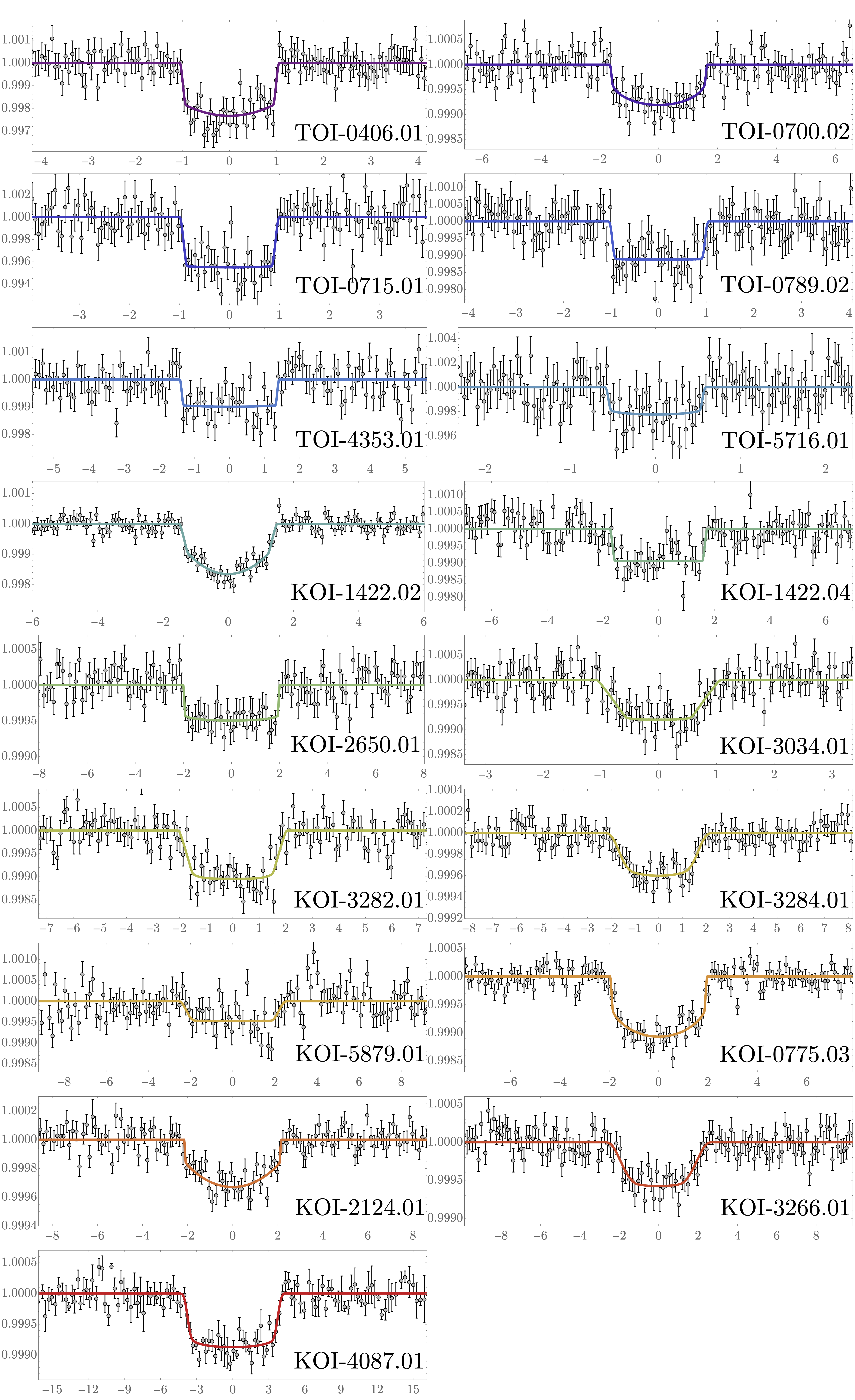}
\caption{\label{fig:lcplots}
\textbf{Maximum likelihood transit fits to the light curves of the 17 Earth
proxies.}
Light curves have been folded upon the maximum likelihood ephemeris parameters
and then been binned to approximately 30 points within the full transit
duration. In each panel, the $x$-axis represents (folded) time in hours from
the moment of inferior conjunction, and the $y$-axis is relative intensity
(dimensionless).
}
\end{sfigure}

\begin{stable}
\caption{
One-sigma credible intervals for the planetary radius
and instellation of the 17 Earth proxies used in our work. The final
column sums the fraction of each object's joint posterior samples
within our pre-defined ``Earth-proxy'' criteria.
} 
\centering 
\begin{tabular}{l l l l}
\hline\hline 
KOI/TOI &
$R_P$\,[$R_{\oplus}$] &
$S_P$\,[$S_{\oplus}$] &
Suitability \\
\hline 
TOI-0406.01 & $1.938_{-0.086}^{+0.094}$ & $2.67_{-0.47}^{+0.55}$ & 74.3\% \\
TOI-0700.02 & $1.117_{-0.069}^{+0.099}$ & $0.77_{-0.06}^{+0.06}$ & ${>}99.99$\% \\
TOI-0715.01 & $1.733_{-0.094}^{+0.100}$ & $0.75_{-0.14}^{+0.16}$ & 98.4\% \\
TOI-0789.02 & $1.423_{-0.106}^{+0.131}$ & $3.03_{-0.52}^{+0.61}$ & 93.5\% \\
TOI-4353.01 & $1.383_{-0.115}^{+0.122}$ & $0.44_{-0.07}^{+0.08}$ & 99.6\% \\
TOI-5716.01 & $1.103_{-0.117}^{+0.129}$ & $3.52_{-0.62}^{+0.73}$ & 75.4\% \\
KOI-1422.02 & $1.880_{-0.214}^{+0.229}$ & $3.42_{-0.75}^{+1.85}$ & 54.9\% \\
KOI-1422.04 & $1.516_{-0.179}^{+0.185}$ & $0.59_{-0.16}^{+0.54}$ & 98.1\% \\
KOI-2650.01 & $1.395_{-0.088}^{+0.097}$ & $3.29_{-0.65}^{+2.60}$ & 67.5\% \\
KOI-3034.01 & $1.737_{-0.129}^{+0.158}$ & $0.90_{-0.23}^{+0.83}$ & 90.9\% \\
KOI-3282.01 & $1.923_{-0.107}^{+0.132}$ & $1.31_{-0.27}^{+1.18}$ & 73.2\% \\
KOI-3284.01 & $1.320_{-0.100}^{+0.149}$ & $2.49_{-0.69}^{+3.25}$ & 75.6\% \\
KOI-5879.01 & $1.430_{-0.198}^{+0.457}$ & $0.59_{-0.52}^{+0.70}$ & 67.0\% \\
KOI-0775.03 & $1.987_{-0.110}^{+0.206}$ & $3.18_{-0.67}^{+2.99}$ & 51.5\% \\
KOI-2124.01 & $1.219_{-0.078}^{+0.128}$ & $3.88_{-0.80}^{+3.25}$ & 53.4\% \\
KOI-3266.01 & $1.624_{-0.110}^{+0.170}$ & $2.17_{-0.54}^{+2.63}$ & 80.0\% \\
KOI-4087.01 & $1.924_{-0.090}^{+0.118}$ & $1.63_{-0.32}^{+1.49}$ & 74.7\% \\ [0.5ex]
\hline\hline 
\end{tabular}
\label{tab:suitability} 
\end{stable}

\begin{sfigure}
\centering
\includegraphics[angle=0, width=6.0cm]{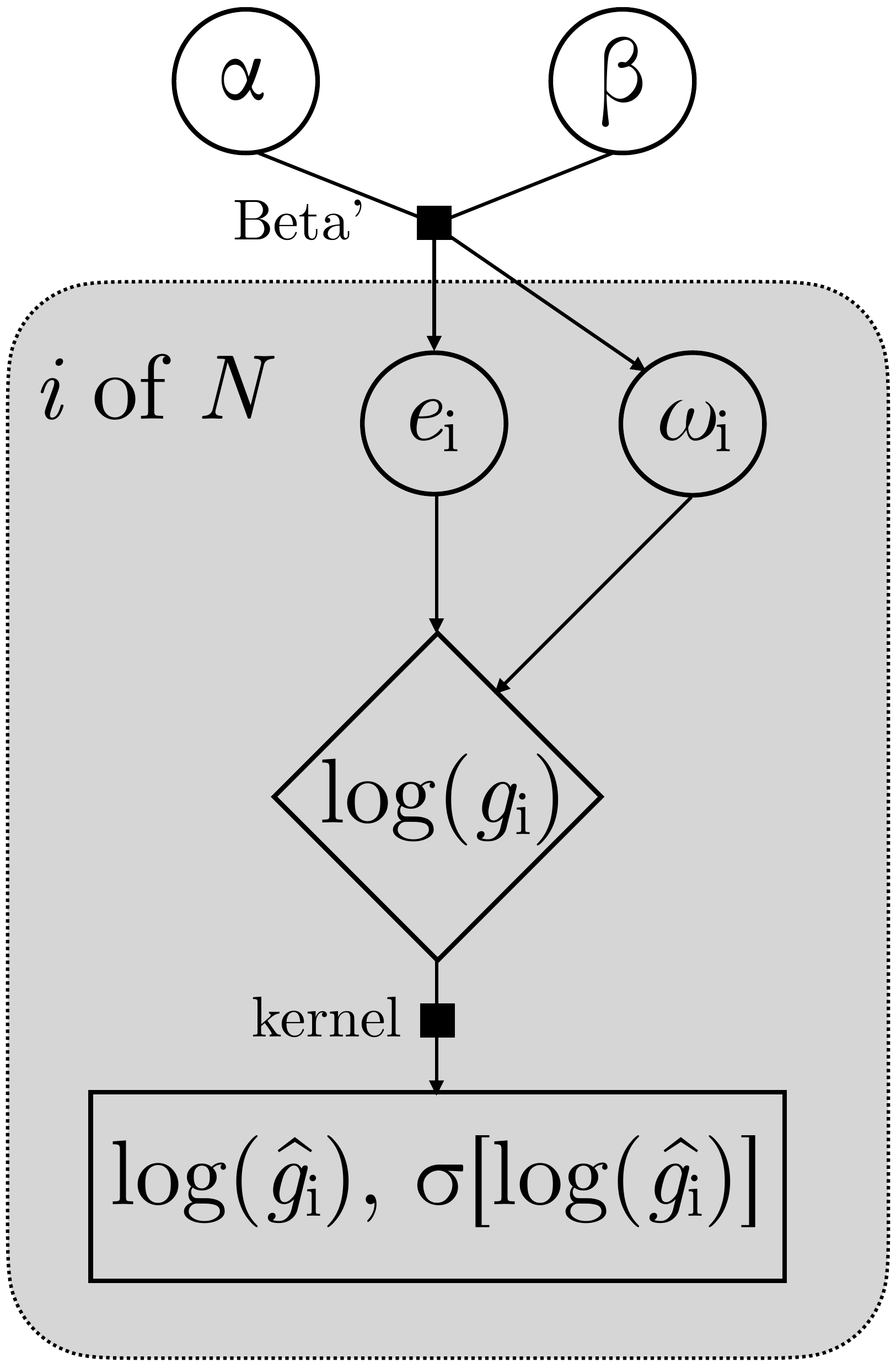}
\caption{\label{fig:graphical_hbm}
\textbf{Graphical model of the HBM used in this work, here in the example
case of a Beta distribution.}
Circles denote free parameters, diamonds intermediate results,
and squares observational data.
All entities on the plate are duplicated $N$
times (where $N$ is the number of planets in our sample). The hyper-parameters
sit at the top of the model, which inform $e_i$ and $\omega_i$ via the
their \textit{a-priori} relation marked with a small black square. That
relation accounts for the transit bias effect, hence why we annotate it is
Beta' rather than strictly Beta. These then combine to produce a $\log\gamma$
value, which is compared to the observed $\log\gamma$ values denoted with a
$\hat{}$ symbol. Here, the relation is not parametric but drawn from a
kernel density estimator.
}
\end{sfigure}

\begin{sfigure}
\centering
\includegraphics[angle=0, width=16.0cm]{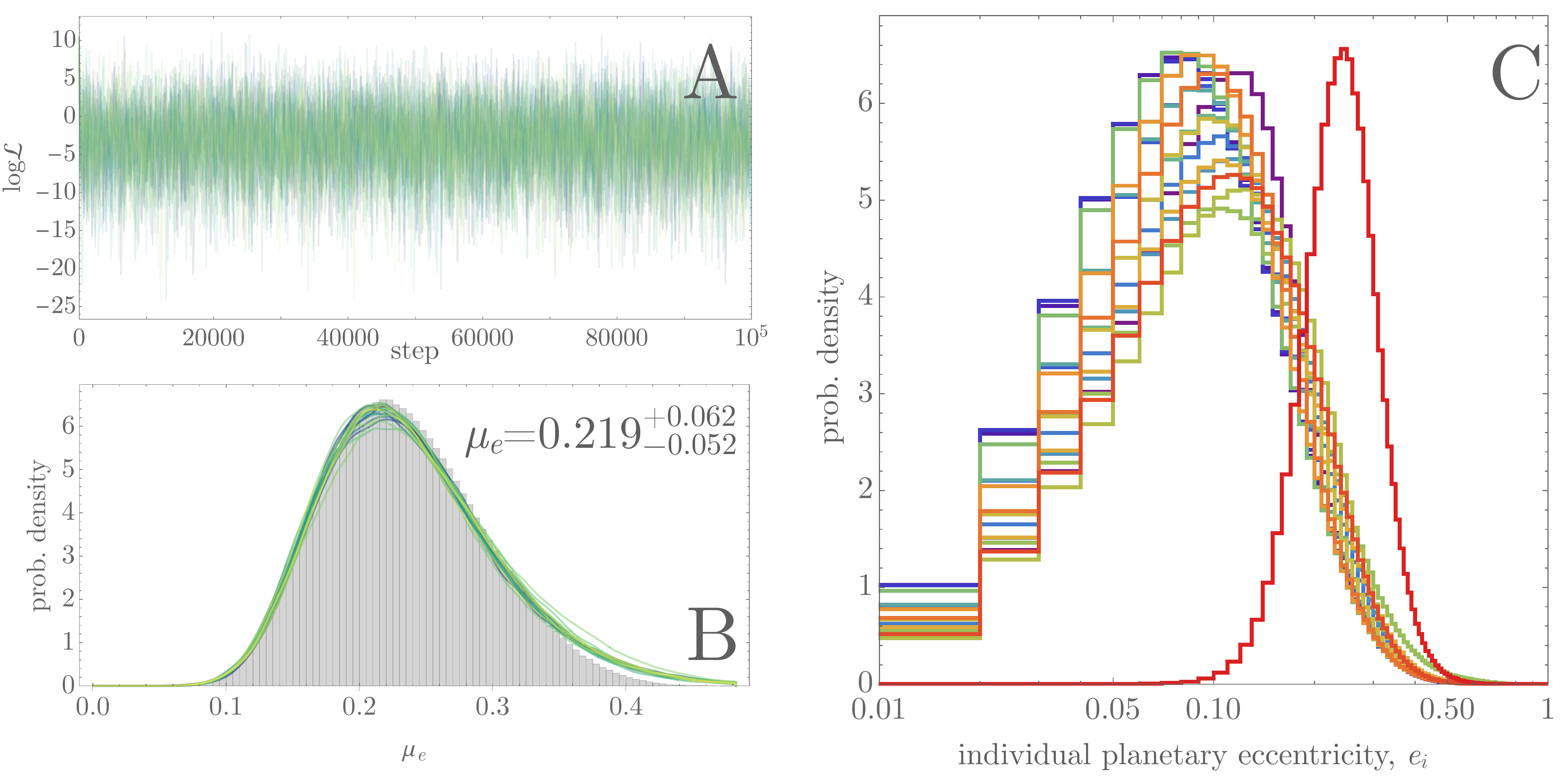}
\caption{\label{fig:tremplot}
\textbf{Same as Figure~\ref{fig:expdist} but for the Tremaine model
(including KOI-4087.01).}
}
\end{sfigure}

\begin{sfigure}
\centering
\includegraphics[angle=0, width=16.0cm]{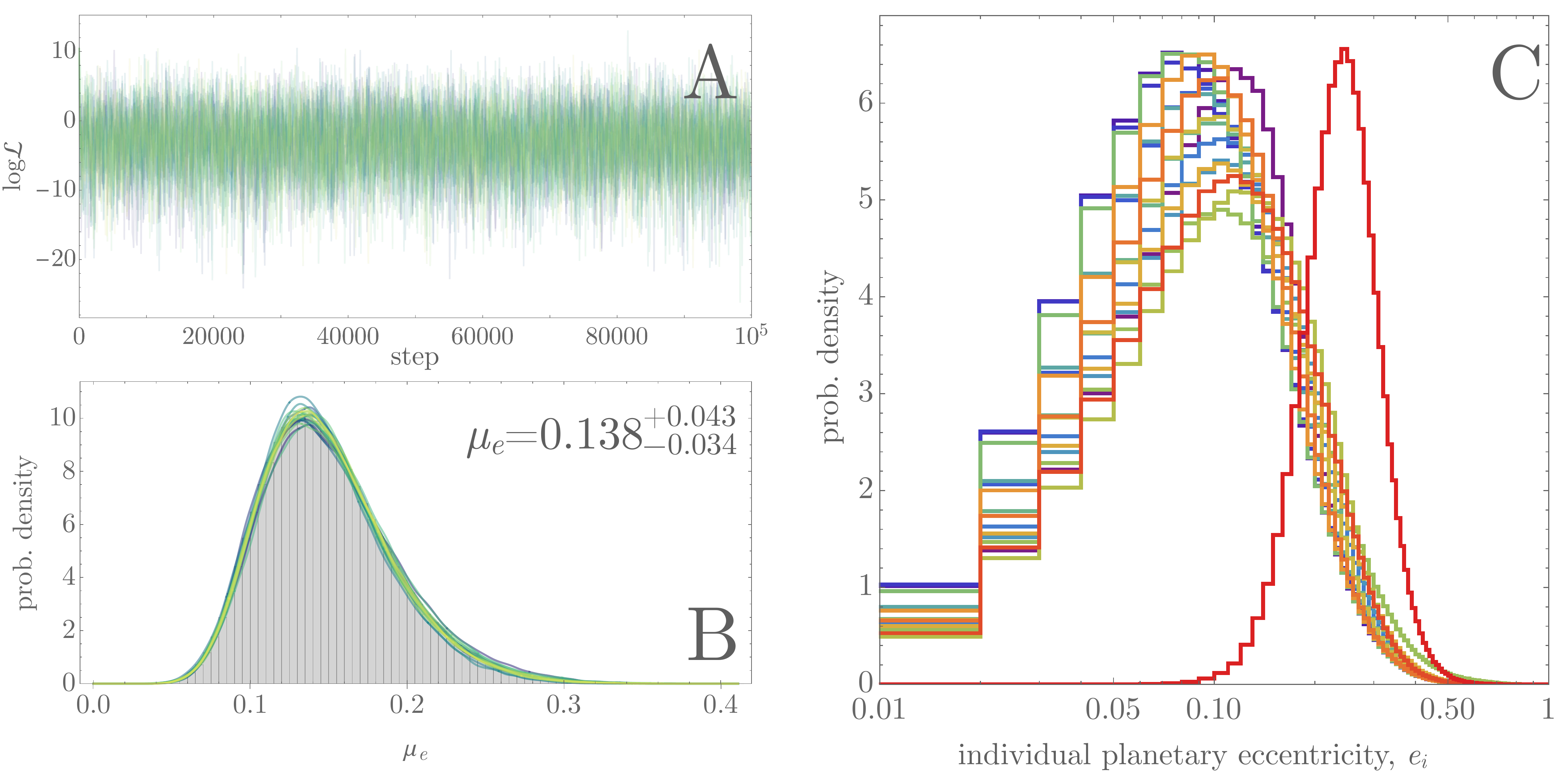}
\caption{\label{fig:raylplot}
\textbf{Same as Figure~\ref{fig:expdist} but for the Rayleigh model
(including KOI-4087.01).}
}
\end{sfigure}

\begin{sfigure}
\centering
\includegraphics[angle=0, width=16.0cm]{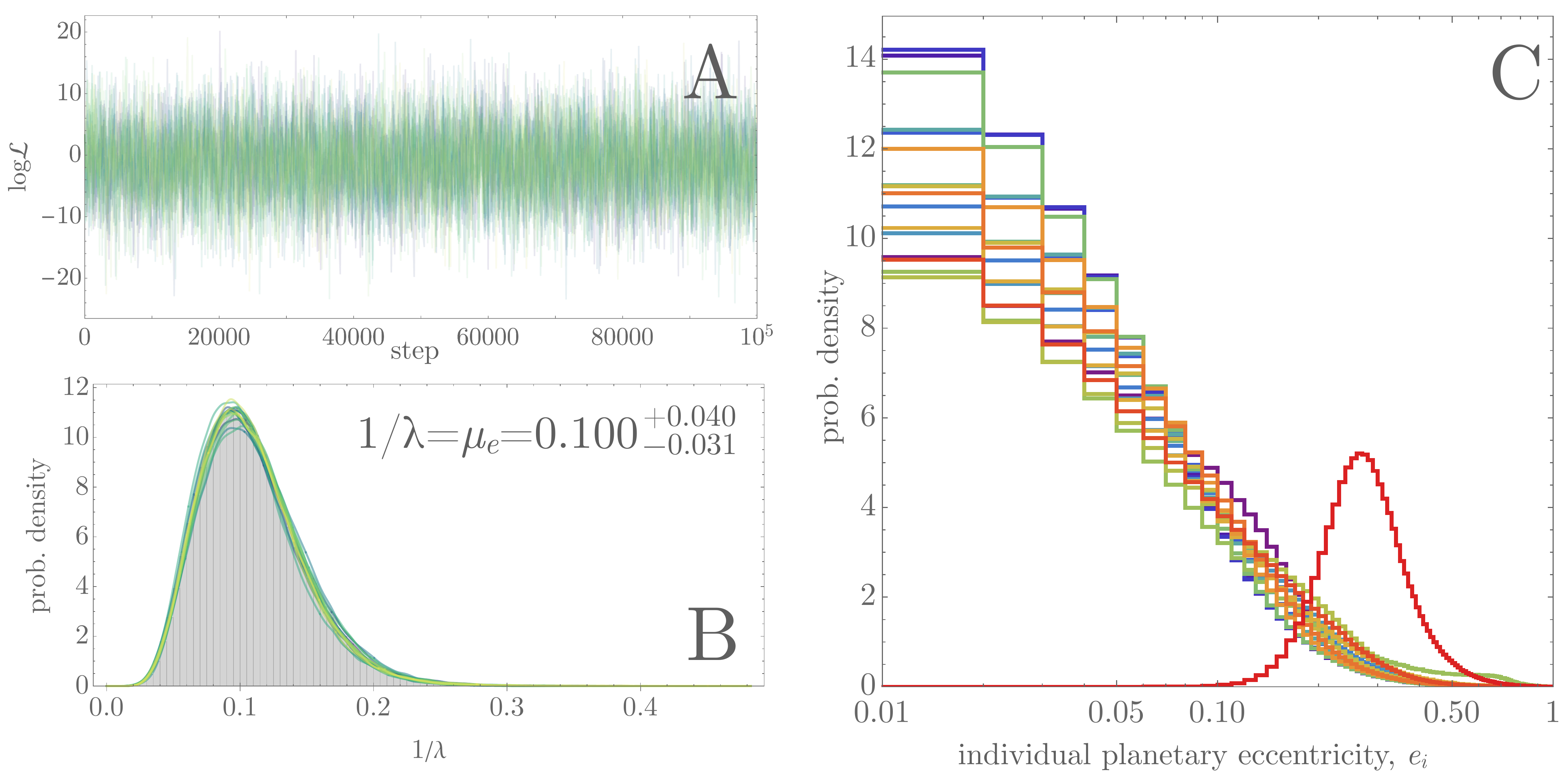}
\caption{\label{fig:expnplot}
\textbf{Same as Figure~\ref{fig:expdist} but for the exponential model
(including KOI-4087.01).}
}
\end{sfigure}

\begin{sfigure}
\centering
\includegraphics[angle=0, width=16.0cm]{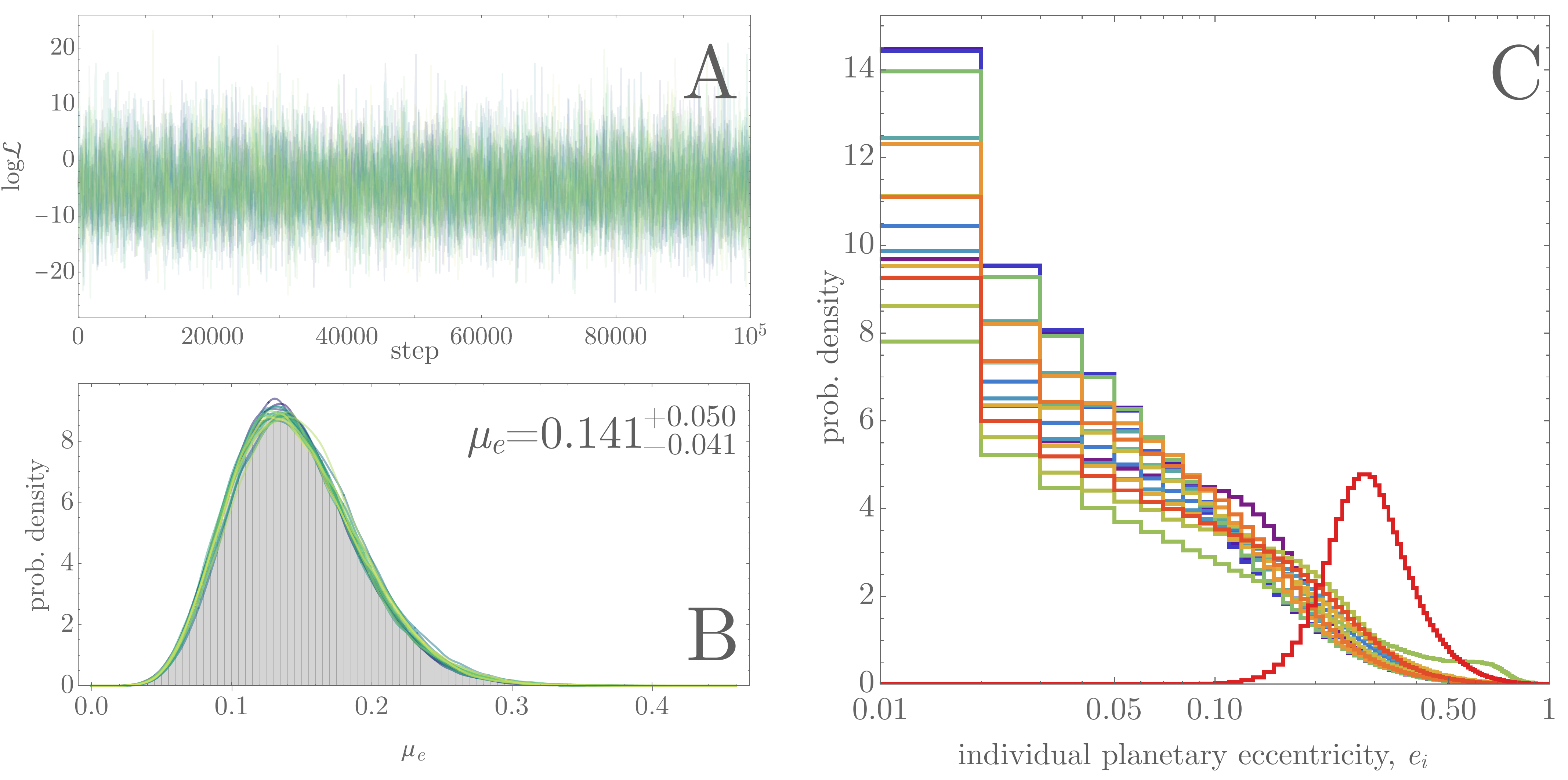}
\caption{\label{fig:betaplot}
\textbf{Same as Figure~\ref{fig:expdist} but for the Beta model
(including KOI-4087.01).}
}
\end{sfigure}

\begin{sfigure}
\centering
\includegraphics[angle=0, width=16.0cm]{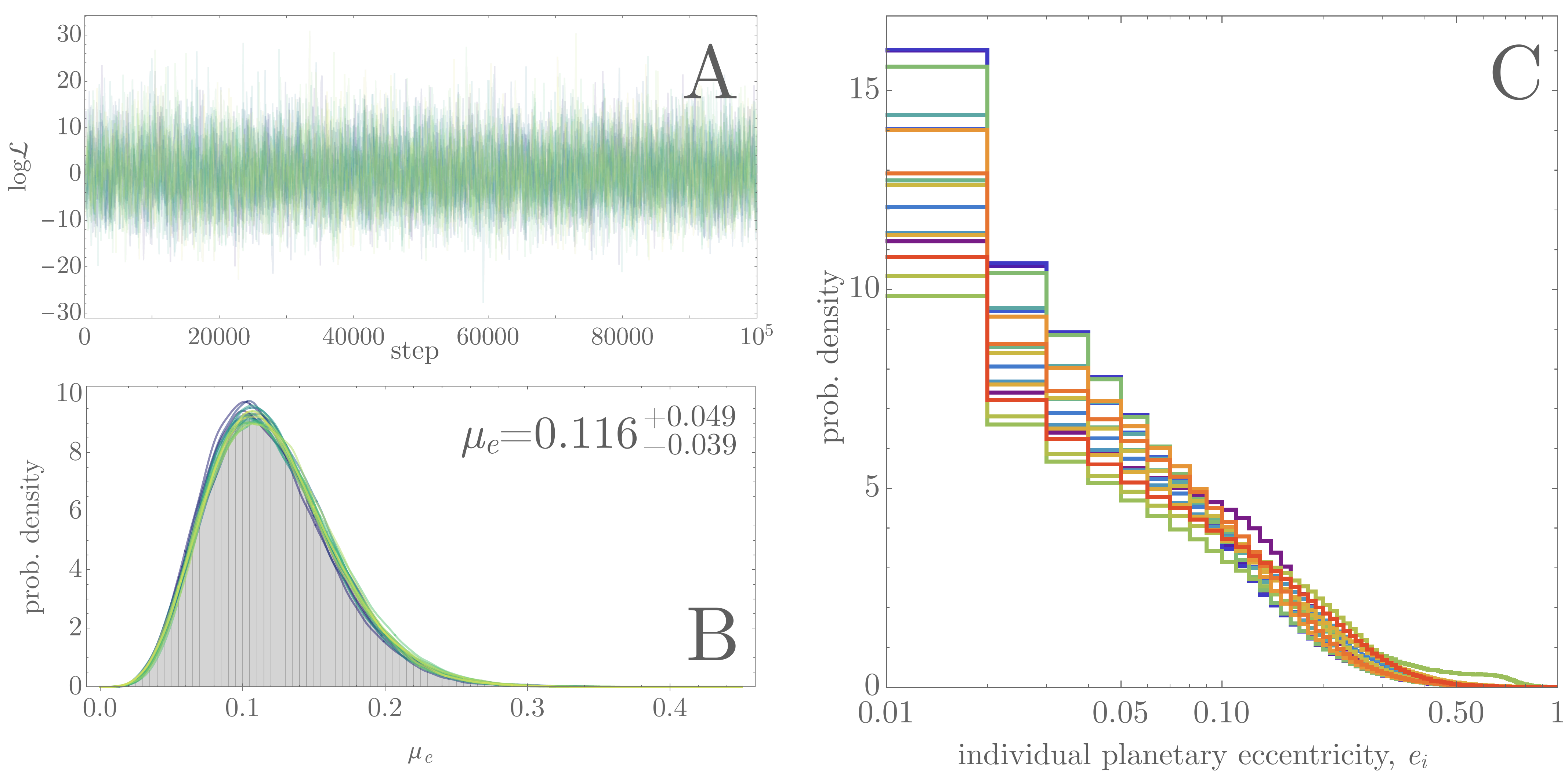}
\caption{\label{fig:betaXplot}
\textbf{Same as Figure~\ref{fig:expdist} but for the Beta model
(excluding KOI-4087.01).}
}
\end{sfigure}

\begin{sfigure}
\centering
\includegraphics[angle=0, width=16.0cm]{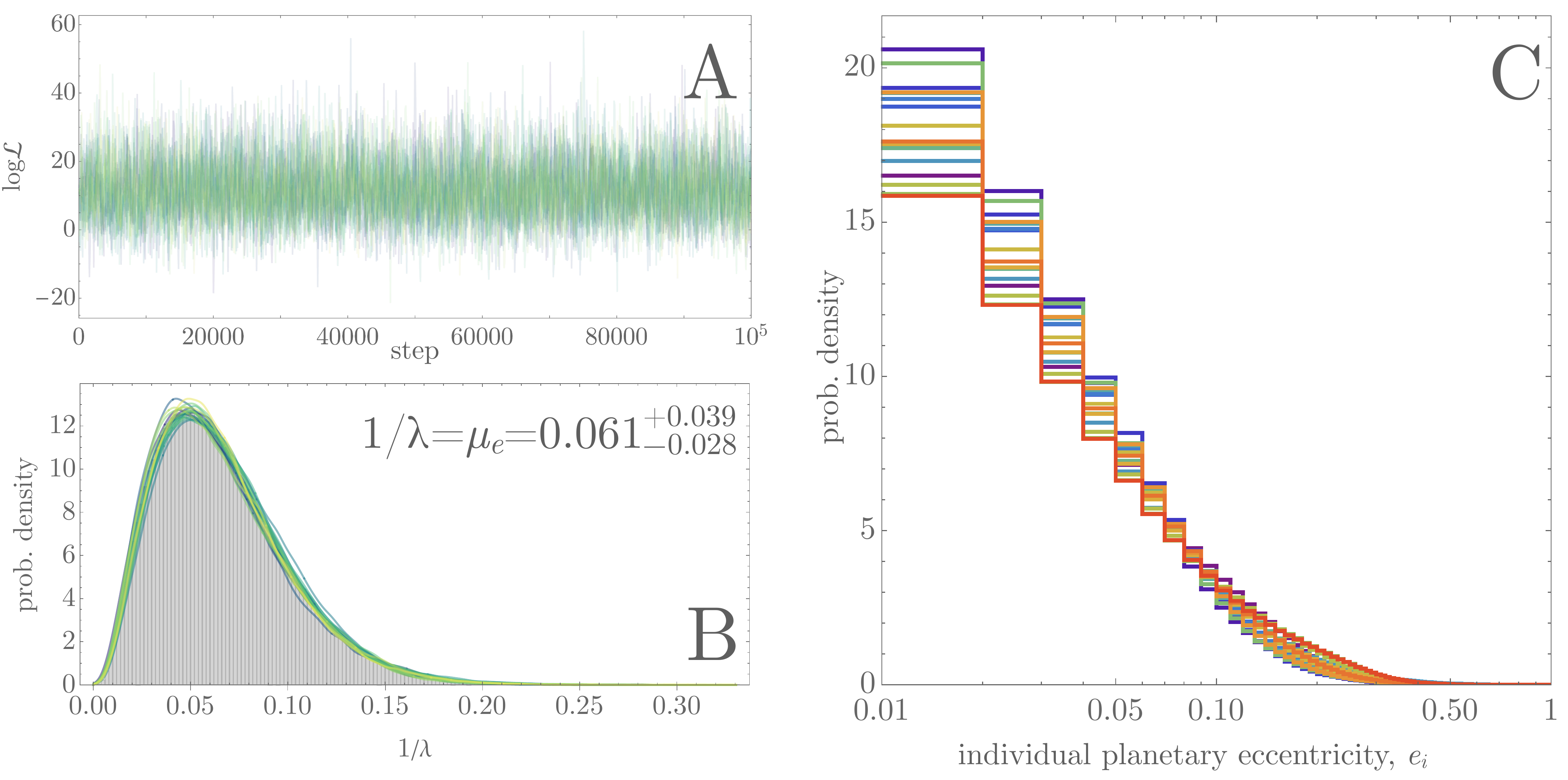}
\caption{\label{fig:expnXCplot}
\textbf{Same as Figure~\ref{fig:expdist} but for the exponential model
(excluding KOI-4087.01) and using the fake circular orbit planets.}
}
\end{sfigure}

\begin{sfigure}
\centering
\includegraphics[angle=0, width=16.0cm]{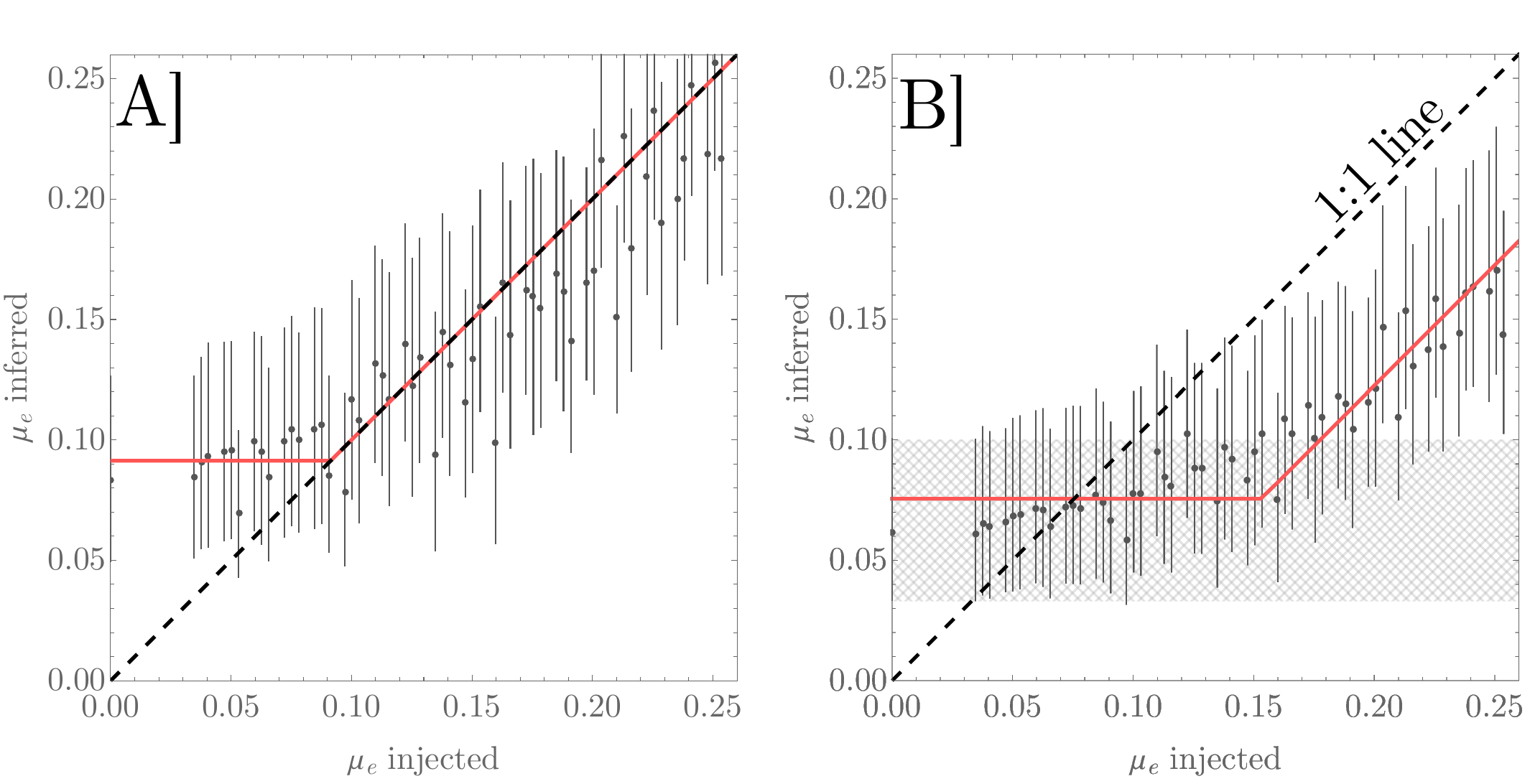}
\caption{\label{fig:hotfakes}
\textbf{Injection-recovery results for eccentric populations.}
A] Each point is an inferred mean eccentricity for an injected
eccentric population of 17 Earth proxies, with the same observational
uncertainties on $\log\gamma$ as the real data. Here, both the
injected and inferred eccentricity distributions are the Rayleigh
distribution. Red line shows a best-fitting piecewise function.
B] Same as A, but the inference model is mis-specified,
specifically it is now an exponential.
}
\end{sfigure}

\begin{sfigure}
\centering
\includegraphics[angle=0, width=16.0cm]{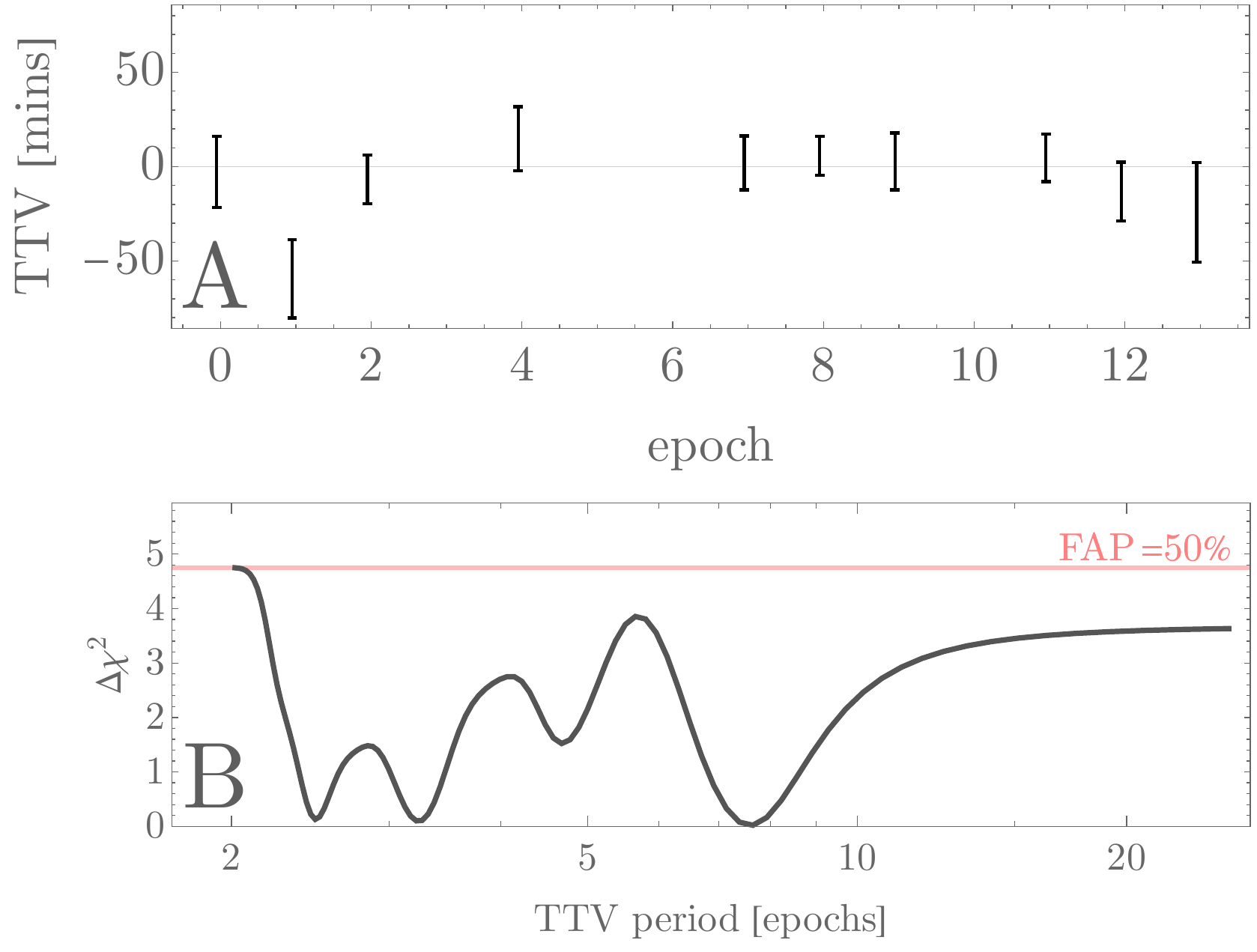}
\caption{\label{fig:4087_ttvs}
\textbf{Transit timing variations (TTVs) for KOI-4087.01.}
Top: TTV diagram showing the 1-$\sigma$ credible intervals for the
10 transit times, after subtracting the best fitting linear ephemeris.
Bottom: Periodogram of the TTVs and the FAP$=50$\% threshold, which
highlights the absence of any significant TTVs here.
}
\end{sfigure}

\end{document}